\documentclass[
  floats,
  floatfix,
  prd,
  twocolumn,
  reprint,
  superscriptaddress,
  nofootinbib,
  longbibliography,
]{revtex4-2}
\usepackage{amsfonts,amssymb,amsmath}
\usepackage[caption=false]{subfig} %
\usepackage{graphicx, hyperref, tabularx, dcolumn}

\usepackage{color}
\usepackage[dvipsnames]{xcolor}
\usepackage[normalem]{ulem}
\usepackage{orcidlink}
\usepackage{tensor, comment}
\usepackage{enumitem}
\usepackage{tensor}

\newcommand{\mach}{\mathcal{M}}
\newcommand{\reffig}[1]{Figure~\ref{#1}}
\newcommand{\refsec}[1]{Sec.~\ref{#1}}
\newcommand{\e}[1]{\times 10^{#1}}

\newcommand{\athenak}{\textsc{AthenaK}}

\newcommand{\caltech}{\affiliation{Theoretical Astrophysics 350-17, California
Institute of Technology, Pasadena, California 91125, USA}}
\newcommand{\burke}{\affiliation{Walter Burke Institute for Theoretical Physics,
California Institute of Technology, Pasadena, California 91125, USA}}

\begin{document}

\author{Yoonsoo Kim \orcidlink{0000-0002-4305-6026}} \email{ykim7@caltech.edu}
\caltech
\affiliation{Department of Physics, California Institute of Technology, Pasadena, California 91125, USA}

\author{Elias R. Most \orcidlink{0000-0002-0491-1210}} \caltech \burke

\title{{General relativistic magnetized Bondi-Hoyle-Lyttleton accretion with
a spin-field misalignment: jet nutation, polarity reversals, and Magnus drag}}

\date{\today}

\begin{abstract}
    The dynamics of a black hole traveling through a plasma -- a general
    relativistic extension of the classic Bondi-Hoyle-Lyttleton (BHL) accretion
    problem -- can be related to a variety of astrophysical contexts, including
    the aftermath of binary black hole mergers in gaseous environments. We
    perform three-dimensional general relativistic magnetohydrodynamics
    simulations of BHL accretion {onto a spinning black hole when magnetic
    field of the incoming wind is inclined} to the spin axis of the black hole.
    Irrespective of inclination but dependent on the wind speed, we find that
    the accretion flow onto the black hole can become magnetically arrested,
    launching an intermittent jet {whose formation is assisted by a
    turbulent dynamo-like process in the inner disk}. The upstream ram pressure
    of the wind bends the jet, and confines the angular extent into which the
    magnetic flux tubes ejected from quasi-periodic eruptions are released.
    Recoil from magnetic flux eruptions drives strong oscillations in the
    {inner accretion disk}, resulting in jet nutation at the outer radii and
    occasionally ripping off the inner part of the accretion disk. {When the
    incoming magnetic field is perpendicular to the spin axis of the black hole,
    we find that the magnetic polarity of the jets can undergo a stochastic
    reversal.} In addition to dynamical friction, the black hole experiences a
    perpendicular drag force analogous to the Magnus effect. Qualitative effects
    of the incoming magnetic field orientation, the strength of the
    magnetization, and the incoming wind speed are investigated as well.
\end{abstract}

\maketitle

\section{Introduction}
\label{sec:intro}

\subsection{BBH merger remnant in AGN disk}

The merger of a binary black hole (BBH)
\cite{LIGOScientific:2016aoc,LIGOScientific:2018mvr,LIGOScientific:2020ibl,KAGRA:2021vkt}
within the accretion disk of an active galactic nucleus (AGN)
\cite{Stone:2016wzz,Tagawa:2019osr,Grobner:2020drr,Ishibashi:2020zzy,McKernan:2021nwk,Ford:2021kcw,Kaaz:2023a}
is thought to be one of major channels of the observed BBH mergers, but is also
an interesting astrophysical scenario within the context of multi-messenger
astronomy. If asymmetry is present in a black hole binary, the resulting
anisotropic emission of gravitational waves from the merger can impart a recoil
onto the post-merger remnant black hole (BH)
\cite{Gonzalez:2006md,Campanelli:2007ew}. Since the kicked remnant will be
moving through a gas-rich environment, a luminous accretion flow onto or a
relativistic jets from the BH may give rise to an observable post-merger
electromagnetic signal \cite{Graham:2020gwr,Chen:2023xrm}.

The gaseous environment of the AGN disk can affect the long-term evolution of an
embedded BBH, often putting constraints on its orbital configuration. Newtonian
studies suggest that orbital and spin axes of a BBH embedded in a gaseous
environment align over time \cite{Bogdanovic:2007hp,ColemanMiller:2013jrk}, and
the orbit is driven to be aligned with the AGN disk plane as well
\cite{Dittmann:2023sha}. These findings, put together, indicate that both the
orbital angular momentum and spin of the BBH are likely aligned with the AGN
disk. Unless anisotropic gravitational radiation induces a significant torque on
the system, the remnant would retain its prior spin direction.

Recent large-scale cosmological simulations revealed that the magnetic field of
AGN disks are predominantly toroidal (parallel to the disk plane)
\cite{Hopkins2024a,Hopkins2024b}; {see also Ref. \cite{Gaburov:2012jd} for
an earlier work.} The presence of mixed poloidal-toroidal configurations is also
in line with simulations of magnetized circumbinary disks around BBHs
\cite{Most:2024qus,Most:2024onq} as well. {Overall, this} motivates a
theoretical investigation on a recoiled black hole flying through a plasma
{embedded with a magnetic field misaligned with} the spin of the black hole.

\subsection{Bondi-Hoyle-Lyttleton accretion}
\label{sec:bhl intro}

Bondi-Hoyle-Lyttleton (BHL) accretion \citep{Bondi1952,Hoyle1939} is a classic
problem in astrophysics involving a gravitational accretor traveling through a
uniform fluid. Despite being highly simplified, it can be applied to a wide
range of astrophysical systems including common envelope phases of binary star
evolution
\cite{MacLeod2015,MacLeod2017,Murguia-Berthier:2017bdf,Lopez-Camara:2020pot},
wind-fed X-ray binaries \cite{ElMellah:2017ncw}, star clusters
\cite{Kaaz:2019wdi} or protoplanetary disks \cite{Moeckel:2009sq}. Owing to its
astrophysical significance, a large volume of analytical and numerical studies
exist in the literature on Newtonian \cite[see ][for a
review]{Edgar:2004mk,Foglizzo:2005in} and relativistic
\cite{Petrich:1988zz,Petrich1989,Tejeda:2019lie,Font:1998,Font1998b,Font:1999,Donmez:2011,Donmez2012,Zanotti:2011mb,Penner:2013,Koyuncu:2014nga,Lora-Clavijo:2013,Lora-Clavijo:2015,Blakely2015,Cruz-Osorio:2012anr,Cruz-Osorio:2016abh,Cruz-Osorio:2020dja,Penner:2011,Gracia-Linares:2015woa,Kaaz:2023b,Gracia-Linares:2023yxw}
regimes.

Basic physical scales associated with BHL accretion are the accretion
radius\footnote{An alternate definition of the accretion radius exists in the
literature
\begin{equation*}
    R_a = \frac{2GM}{c_{s,\infty}^2 + v_\infty^2},
\end{equation*}
where $c_{s,\infty}$ is the asymptotic sound speed of the incoming fluid. We
adopt the definition \eqref{eq:accretion radius} throughout this paper.}
\begin{equation} \label{eq:accretion radius}
    R_a = \frac{2GM}{v_\infty^2} \, ,
\end{equation}
the accretion timescale
\begin{equation} \label{eq:accretion timescale}
    \tau_a = \frac{R_a}{v_\infty} = \frac{2GM}{v_\infty^3} \, ,
\end{equation}
and the Bondi-Hoyle-Lyttleton mass accretion rate
\begin{equation} \label{eq:bhl accretion rate}
    \dot{M}_\text{BHL} = \pi R_a^2 \rho_\infty v_\infty
        = \frac{4\pi G^2 M^2 \rho_\infty}{v_\infty^3} \, ,
\end{equation}
where $G$ is the gravitational constant, $M$ is the mass of the accreting
object, $v_\infty$ is the asymptotic relative velocity, and $\rho_\infty$ is the
asymptotic mass density of the fluid.

A major challenge in computational approaches to this problem is its inherent
multi-scale nature, namely simultaneously resolving the size of the accreting
object $r_0$ and the accretion radius $R_a$ on a single numerical grid. For
example, a black hole with mass $M$ has $r_0 \approx r_g$ where
\begin{equation}
    r_g = \frac{GM}{c^2}
\end{equation}
is the gravitational radius of the black hole. The ratio between the two length
scales is
\begin{equation} \label{eq:length scale separation}
    \frac{R_a}{r_g} \sim \left( \frac{v_\infty}{c} \right)^{-2}.
\end{equation}
Also, time integration needs to be performed at least several times of $\tau_a$
to reach a steady state, which is longer than the dynamical timescale associated
with the black hole by a factor of
\begin{equation} \label{eq:time scale separation}
    \frac{\tau_a}{(r_g/c)} \sim \left( \frac{v_\infty}{c} \right)^{-3}.
\end{equation}

A large separation in both length \eqref{eq:length scale separation} and time
\eqref{eq:time scale separation} scales, which is especially severe for a
compact accretor such as a black hole, rapidly increases the computational cost
for realistic values of $v_\infty$. As a result, many studies are often forced
to assume an unrealistically fast velocity of the BH relative to the fluid. Due
to its high computational cost, most numerical studies on general relativistic
BHL accretion have considered hydrodynamic flows on either 2D planar or 3D
axisymmetric geometry
\cite{Font:1998,Font1998b,Font:1999,Donmez:2011,Donmez2012,Zanotti:2011mb,Penner:2013,Koyuncu:2014nga,Lora-Clavijo:2013,Lora-Clavijo:2015,Blakely2015,Cruz-Osorio:2012anr,Cruz-Osorio:2016abh,Cruz-Osorio:2020dja},
However, inclusion of magnetic fields can dramatically alter the flow
morphology, and a restrictive nature of the assumed spatial symmetry might fail
to fully capture multi-dimensional effects. Following the first study {of 2D
{magnetohydrodynamic} (MHD)} \cite{Penner:2011} and 3D hydrodynamic
\cite{Gracia-Linares:2015woa} flows, the first simulations of {general
relativistic magnetohydrodynamics (GRMHD)} Bondi-Hoyle-Lyttleton accretion in
full 3D have been carried out only recently
\cite{Kaaz:2023b,Gracia-Linares:2023yxw}. Each of these studies respectively
explored jet launching from the BH \cite{Kaaz:2023b} and the effect of the BH
spin-wind orientation on the shock morphology \cite{Gracia-Linares:2023yxw},
which can only be properly captured in a 3D MHD simulation.

In this paper, we perform {GRMHD} simulations of a spinning black hole
traveling through a fluid {embedded with a magnetic field inclined} to the
spin axis of the BH. The physical scenario is approximated by a relativistic
Bondi-Hoyle-Lyttleton accretion problem with a magnetized wind. We examine
large-scale morphology and temporal evolution of the accretion flow, both of
which are closely related to intermittent jet launching and magnetic flux
eruptions from the BH. The impacts of magnetic field orientation, magnetization,
and the wind speed are assessed by systematically varying simulation parameters.

Another purpose of our numerical experiment is to measure the {outflow}
luminosity (power) and determine the efficiency with which $\dot{M}_{\rm BHL}
c^2$ can be converted into {an energy outflow}. The drag force exerted on
the accreting BH is also measured and its astrophysical implications are
discussed.

This article is organized as follows. In \refsec{sec:methods}, we describe our
numerical setup and methods. We present our results in two steps, focusing on a
specific parameter set first in \refsec{sec:results-fiducial}, before
generalizing it in \refsec{sec:results-parameter-studies}. We present
discussions on the results in \refsec{sec:discussion}, then summarize our main
findings along with limitations and future perspectives in
\refsec{sec:conclusion}.

\section{Methods} \label{sec:methods}

The background spacetime is set to the Kerr metric in (Cartesian) Kerr-Schild
coordinates. The spin of the black hole is aligned with the $\hat{z}$ axis of
the computational domain. The exact form of the spacetime metric in these
coordinates is
\begin{equation} \label{eq:kerr metric in ks coordinates}
\begin{split}
ds^2 & =  - {c^2} dt^2 + dx^2 + dy^2 + dz^2
    + \frac{2{G} Mr^3 {/c^2}}{r^4 + a^2 z^2} \\
    \times  &
     \bigg[ {c\,} dt + \frac{r(x dx + y dy) + a(y dx - x dy)}{r^2 + a^2}
        + \frac{z dz}{r} \bigg]^2  \, ,
\end{split}
\end{equation}
where $M$ is the mass and {$a$} is the spin parameter of the black hole
{(with the unit of length)}. The coordinate variable $r$ is defined as
\begin{equation}
    \frac{x^2 + y^2}{r^2 + a^2} + \frac{z^2}{r^2} = 1 \, .
\end{equation}

We solve the equations of ideal GRMHD, which are given in terms of the rest-mass
density current
\begin{align}
 J^\mu  = \rho u^\mu\,,    
\end{align}
and the stress-energy tensor,
\begin{align}
    T^{\mu\nu} = \left(\rho + e + p + b^2\right) u^\mu u^\nu + \left(P + \frac{b^2}{2}\right) g^{\mu\nu} - b^\mu b^\nu\,,
\end{align}
with its electromagnetic component
\begin{align}
    T^{\mu\nu}_{\rm EM} = b^2 u^\mu u^\nu + \frac{b^2}{2} g^{\mu\nu} - b^\mu b^\nu\,,
\end{align}
where $\rho$ is the rest mass density, $e$ is the internal energy density, $p$
is the pressure, $u^\mu$ is the four-velocity, and $b^\mu$ the comoving magnetic
field of the fluid, with $b^2 = b^\mu b_\mu$. The electromagnetic field is
evolved using the dual field strength tensor,
\begin{align}
    \,^{\ast}\!F^{\mu\nu} = b^\mu u^\nu - u^\mu b^\nu\,. 
\end{align}
The evolution equations are
\begin{align}
    \nabla_\mu J^\mu &=0\,,\\
    \nabla_\mu \,^{\ast}\!F^{\mu\nu} &=0\,,\\
    \nabla_\mu T^{\mu\nu} &=0 \, .
\end{align}

We find it advantageous to define a normal magnetic field as\footnote{This
definition of the magnetic field is not covariant, as it is different by a
factor of $\alpha$ (the lapse function in 3+1 decomposition) from the Eulerian
magnetic field $B^i$ commonly adopted in numerical relativity or relativistic
electrodynamics literature
\cite[{e.g.,}][]{baumgarte_numerical_2010,paschalidis_2013,Komissarov2004}.
However, we adopt the definition \eqref{eq:magnetic field definition} here since
it simplifies the handling of the solenoidal constraint on the magnetic field
as,
\begin{equation*}
    \partial_i \bar{B}^i = 0\, ,
\end{equation*}
in the Kerr-Schild coordinates. }
\begin{align} \label{eq:magnetic field definition}
    \bar{B}^i = \,^{\ast}\!F^{0i}\,. 
\end{align}

We further model the gas dynamics using an ideal fluid equation of state
\begin{align}
    p = e \left(\Gamma -1\right)\,,
\end{align}
where $\Gamma = 5/3$ is the adiabatic index.

\subsection{Numerical setup}

We numerically solve the ideal GRMHD system using the \athenak~code
\cite{Stone2024}, a rewrite of the \textsc{Athena++} \citep{Stone2020} using the
performance portability library Kokkos \citep{kokkos:2022}.

The computational domain $[-40960 r_g, 40960 r_g]^3$ is discretized into a
uniform Cartesian grid with the base grid resolution $256^3$. 13 levels of
static mesh refinement are applied around the coordinate origin, with the
innermost mesh $[-5r_g, 5r_g]^3$ providing a resolution of $\sim$26 grid points
per $r_g$.
Time integration is performed using a second order Runge-Kutta stepper,
piecewise parabolic reconstruction \cite{Colella:1982ee}, an HLLE Riemann solver
\cite{Harten1983,Einfeldt1991}, and a constrained transport algorithm
\cite{Gardiner:2007nc}.
We use the mass density and internal energy floor values $\rho_\text{floor} =
10^{-14}\rho_\infty$, $e_\text{floor} = \rho_\text{floor}c^2/3$, and cap the
maximum Lorentz factor of the fluid to $W_\text{max} = 20$. The drift frame
flooring technique \citep{Ressler2017} is applied to limit the comoving
magnetization to $\sigma_\text{max} = 50$. 

\athenak~can be compiled and run on graphics processing unit (GPU) devices at
scale, providing $\mathcal{O}(10^7)$ cell updates per second per each GPU card
for large scale GRMHD simulations. The simulations presented here overall have
been performed on 132 GPU nodes (792 NVIDIA Volta cards) at OLCF Summit cluster,
costing 900 node hours per $10^4 r_g/c$ integration time on average. The total
computational cost used for all simulations is about 34,000 node hours.

\subsection{Initial data}

\begin{table*}
\centering
\caption{List of the models and parameters considered in this work. Each column
    denotes {the BH rotational (spin) parameter $a$}, accretion radius
    $R_a$, asymptotic fluid incoming speed $v_\infty$, asymptotic Mach number
    $\mach$, accretion time scale $\tau_a$, the plasma beta parameter of the
    incoming fluid $\beta_\infty$, and the magnetic field inclination angle
    $\theta_B$. {The spin of the BH is set to $a=0.9r_g$ for all
    simulations}.}
\label{tab:parameters}
\setlength{\tabcolsep}{8pt}
\begin{tabular}{c|ccccccc|c}
\hline
Label & {$a$} & $R_a$ & $v_\infty/c$ & $\mach$ & $\tau_a$ & $\beta_\infty$ & $\theta_B$
    & Comments \\
& {$[r_g]$} & $[r_g]$ & & & $[r_g/c]$ & & & \\
\hline\hline
$\beta_{10}$-$\theta_{90}$-$R_{200}$ & {0.9} & 200 & 0.1 & 2.0 & 2000 & 10 & 90$^\circ$ & Fiducial setup \\
\hline
$\beta_{10}$-$\theta_{68}$-$R_{200}$ & & 200 & 0.1 & 2.0 & 2000 & 10 & 67.5$^\circ$ & Varying
$\theta_B$ \\
$\beta_{10}$-$\theta_{45}$-$R_{200}$ & & 200 & 0.1 & 2.0 & 2000 & 10 & 45$^\circ$ & \\
$\beta_{10}$-$\theta_{23}$-$R_{200}$ & & 200 & 0.1 & 2.0 & 2000 & 10 & 22.5$^\circ$ & \\
$\beta_{100}$-$\theta_{90}$-$R_{200}$ & & 200 & 0.1 & 2.0 & 2000 & 100 & 90$^\circ$ & Weaker magnetization \\
$\beta_{10}$-$\theta_{90}$-$R_{50}$ & & 50 & 0.2 & 4.0 & 250 & 10 & 90$^\circ$ &
    Faster speed \\
$\beta_{10}$-$\theta_{90}$-$R_{400}$ & & 400 & 0.07 & 1.4 & 5660 & 10 & 90$^\circ$
    & Slower speed \\
\hline
\end{tabular}
\end{table*}

{Over the whole computational domain, matter profile} is initialized with a
uniform rest mass density $\rho_\infty$ and spatial velocity
\begin{equation} \label{eq:init-velocity}
    u^{i'} = \left(- \frac{v_\infty}{\sqrt{1-v_\infty^2/c^2}}, 0, 0\right),
\end{equation}
where $v_\infty$ is the asymptotic incoming speed of the fluid, $c$ is the speed
of light, and $u^{i'}=u^i + \beta^i u^0$ is the normal-frame spatial velocity
which is a primitive variable used in the code.\footnote{See {e.g.,} the
Sec. 4 of Ref. \cite{Stone2020} for the definition of GRMHD primitive variables
used in \textsc{Athena++} {or the Sec. 3 of Ref. \cite{Stone2024} for
\athenak.}}
Given the sound speed $c_{s,\infty}$, the fluid internal energy density
\begin{equation} \label{eq:init-e} 
    e_\infty = \frac{c_{s,\infty}^2 \rho_\infty}{\Gamma(\Gamma - 1 - c_{s,\infty}^2 / c^2)},
\end{equation}
and pressure
\begin{equation} \label{eq:init-p}
    p_\infty = e_\infty (\Gamma - 1),
\end{equation}
can be initialized accordingly.

{The inclination between the magnetic field of the incoming wind and its
velocity can be arbitrary in general. In this paper, in order to narrow down the
parameter space and focus on the misalignment between the BH spin and the
magnetic field, we assume that the incoming magnetic field is perpendicular to
the wind velocity.} We initialize the magnetic field as
\begin{equation} \label{eq:initial b field}
    \bar{B}^i = B_0 (0, \sin \theta_B, \cos \theta_B),
\end{equation}
where $B_0$ is the field strength and $\theta_B$ is the inclination angle
between the BH spin and the magnetization of the incoming fluid.

In the asymptotic limit ($x^i\to\infty$), the magnetic field strength $B_0$ is
related to the magnetization of the fluid $\sigma$ as
\begin{equation} \label{eq:init-sigma}
    \sigma_\infty = \frac{(b^2)_\infty}{(\rho c^2 + e + p)_\infty}
     = \frac{B_0^2 / (1-v_\infty^2/c^2)}{\rho_\infty c^2 + \Gamma e_\infty} \, .
\end{equation}
The plasma $\beta$-parameter and the magnetization $\sigma$ are related via
\begin{equation} \label{eq:init-beta}
    \beta_\infty = \frac{p_\text{gas}}{p_b} = \frac{p_\infty}{(b^2)_\infty/2}
        = \frac{2}{\Gamma} \frac{(c_{s,\infty}/c)^2}{\sigma_\infty} .
\end{equation}

Input parameters of our simulations are the asymptotic fluid mass density
$\rho_\infty$, accretion radius $R_a$, asymptotic sound speed $c_{s,\infty}$,
the plasma parameter $\beta_\infty$, and the magnetic field inclination angle
$\theta_B$. The incoming speed of the fluid $v_\infty$ is computed from
$v_\infty^2 = 2GM/R_a$ and used to initialize the spatial velocity
\eqref{eq:init-velocity}. From the sound speed $c_{s,\infty}$, both internal
energy density and pressure can be initialized using \eqref{eq:init-e} and
\eqref{eq:init-p}. The plasma parameter $\beta_\infty$ determines the
relativistic magnetization $\sigma_\infty$ via \eqref{eq:init-beta}, in turn
giving $B_0$ from \eqref{eq:init-sigma}, which we use to initialize the magnetic
fields as \eqref{eq:initial b field}.
Primitive variables are initialized everywhere in the computational domain with
the values prescribed as above. 

Simulation parameters are listed in Table~\ref{tab:parameters}. We particularly
focus on the cases that the magnetic field of the incoming wind is
{perpendicular or inclined} with respect to the BH spin (see Ref.
\cite{Kaaz:2023b} for the {aligned} magnetic field configuration). For all
simulations we adopt a BH spin {$a/r_g = 0.9$} and the asymptotic sound
speed $c_{s,\infty} = 0.05c$. Our representative, fiducial model assumes
$\beta_\infty=10$ with a {horizontal} orientation of the magnetic field
$\theta_B = 90^\circ$, along with the fluid incoming speed $v_\infty = 0.1c$
corresponding to $R_a = 200 r_g$; this model is labeled as
$\beta_{10}$-$\theta_{90}$-$R_{200}$. To explore the influence of magnetic field
orientation relative to the BH spin, we run the identical setup but varying
$\theta_B = 22.5^\circ, 45^\circ, 67.5^\circ$, each of which is labeled with
$\theta_{23}$, $\theta_{45}$, and $\theta_{68}$. We perform an experiment on the
effect of a weaker magnetization $\beta_\infty = 100$ while keeping other
parameters fixed from the fiducial model. Lastly, we run two more models with a
smaller ($R_a = 50r_g$) and a larger ($R_a = 400r_g$) accretion radius to test
the influence of the fluid incoming speed, also keeping the other parameters
fixed to the same values as the fiducial model.

{ 
\subsection{Boundary conditions}
At the outer boundary of the computational domain facing the $+\hat{x}$
(upstream) direction, we impose a Dirichlet boundary condition injecting a
constant wind profile same as the initial data. At all other sides of the
domain boundary, primitive variables at the outermost grid points are copied
into the ghost zones to impose a free streaming boundary condition.}

\subsection{Analysis} \label{sec:analysis}

In addition to the magnetohydrodynamics variables evolved on the grid, we
compute following integral quantities from simulation data in order to monitor
time evolution of the system:
\begin{itemize}[leftmargin=3ex, labelsep=1ex]
\item Mass accretion rate
    \begin{equation}
        \dot{M} = \oint (-\rho u^r) \sqrt{-g} \, d\theta d\phi ,
    \end{equation}
    where $\rho$ is the rest mass density and $u^r$ is the radial component of
    the four-velocity. Note the minus sign in the integrand, which makes a
    positive value indicate mass inflow.
\item Total energy (in-)flux
    \begin{equation}
        \dot{E} = \oint \tensor{T}{^r_t} \sqrt{-g} \, d\theta d\phi \,.
    \end{equation}
\item Angular momentum (out-)flux
    \begin{equation}
        \dot{J} = \oint \tensor{T}{^r_\phi} \sqrt{-g} \, d\theta d\phi .
    \end{equation}
\item Total magnetic fluxes threading the horizon
    \begin{equation}
        \Phi_\text{BH} = \frac{1}{2} \oint_{\rm BH} |\bar{B}^r| \sqrt{-g} \, d\theta d\phi .
    \end{equation}
    For our discussions presented here and in the following sections, we will
    often refer to the dimensionless total magnetic fluxes threading the BH
    horizon
    \begin{equation}
        \phi_\mathrm{BH} \equiv \frac{\Phi_\text{BH}}{\sqrt{\dot{M} r_g^2 c}}\,.
    \end{equation}
    The dimensionless magnetic flux can be used as an indicator for the
    accretion state, showing that for $\phi_\mathrm{BH} \gtrsim 20$ jets can be
    launched \cite{Tchekhovskoy:2011zx}. For larger values of
    $\phi_\mathrm{BH}$, the accretion flow will become fully magnetically
    arrested
    \cite[{e.g.,}][]{Tchekhovskoy:2011zx,McKinney:2012vh,White:2019csp,Chatterjee:2022mxg}.
\item Drag force exerted on the BH
    \begin{equation} \label{eq:total drag force}
        F^i = - F_\text{mom}^i + F_\text{grav}^i ,
    \end{equation}
    where $F_\text{mom}^i$ is the total momentum (out-)flux through a spherical
    surface
    \begin{equation}
        F_\text{mom}^i = \oint \tensor{T}{^r_i} \sqrt{-g} \, d\theta d\phi ,
    \end{equation}
    and $F_\text{grav}^i$ is the gravitational drag computed with a modified
    Newtonian formula \cite{Cruz-Osorio:2020dja,Kaaz:2023b}
    \begin{equation} \label{eq:gravitational drag}
        F_\text{grav}^i = \int \rho \frac{x^i}{r^3} dV .
    \end{equation}
    A minus sign on the right hand side of Eq.~\eqref{eq:total drag force}
    accounts for that the momentum loss in a closed volume results in a reaction
    force to the opposite direction. For example, in our simulation setup in
    which BH travels to $+x$ direction, $F_\text{mom}^x > 0$ corresponds to the
    deceleration of the BH with respect to the ambient medium, where
    $F_\text{grav}^x > 0$ corresponds to the acceleration.
\end{itemize}

The flow in the vicinity of the BH is strongly magnetized in our simulations,
frequently triggering density and energy floors on troubled grid cells. Since
the artificial injection of mass and energy density from numerical flooring can
contaminate some of the diagnostics introduced above, we perform surface
integrals around the BH at a slightly larger radius than the outer event horizon
\cite[{e.g.,}][]{Kaaz:2023b}. In our analysis, the horizon magnetic fluxes
$\Phi_\text{BH}$ is integrated over the outer event horizon whereas $\dot{M}$,
$\dot{E}$, $\dot{J}$, and $F_\text{mom}^i$ are extracted at $r=3r_g$. The
gravitational drag $F_\text{grav}^i$ is integrated over the whole volume of the
computational domain excluding $r < 3r_g$ in the same regard. We find this
approach has no significant effect our diagnostics, see
Appendix~\ref{sec:r-dependence} for a detailed analysis.

The turbulent nature of the accretion flow leads to high-frequency fluctuations
of the time series data. For improved readability, we have averaged all time
series data over a sliding time window of $100 r_g/c$, unless otherwise stated.  

Computations are performed in a scale-free manner by setting
$c=GM=\rho_\infty=1$. We list unit conversions in Appendix \ref{app:units}.

\section{Fiducial model: Jet launching from horizontally magnetized wind}
\label{sec:results-fiducial}

In this section, we analyze the results from the baseline model
$\beta_{10}$-$\theta_{90}$-$R_{200}$ in detail to highlight several key
phenomena observed in the simulation. This model features a wind speed
$v_\infty=0.1c$ and a magnetic field {perpendicular (horizontal)} to the
black hole spin axis, with its strength set by the plasma parameter $\beta=10$.

\subsection{Overview} \label{sec:fiducial-overview}

\begin{figure*}
\centering
\includegraphics[width=\linewidth]{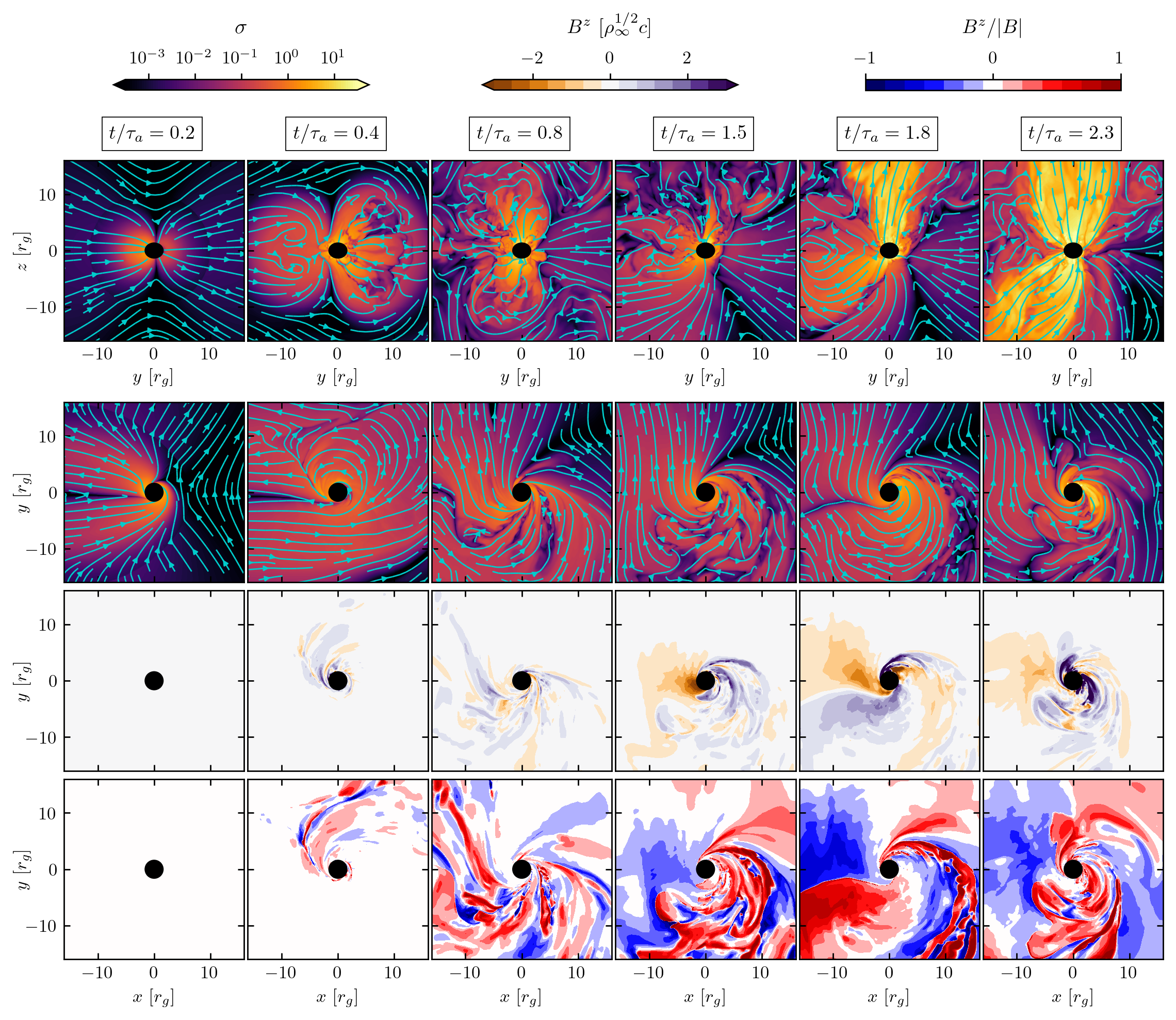}
\caption{{Initial jet launching process in the
    $\beta_{10}$-$\theta_{90}$-$R_{200}$ simulation. Shown here are in-plane
    magnetic field lines (cyan solid lines), magnetization $\sigma$, vertical
    magnetic field $B^z$, and its normalized value $B^z/|B|$. Each column
    displays physical quantities on the vertical ($yz$, the face-on direction
    with respect to the incoming wind) and the equatorial ($xy$) plane at each
    simulation time.}}
\label{fig:jet-buildup}
\end{figure*}

\begin{figure*}
\centering
\includegraphics[width=0.97\linewidth]{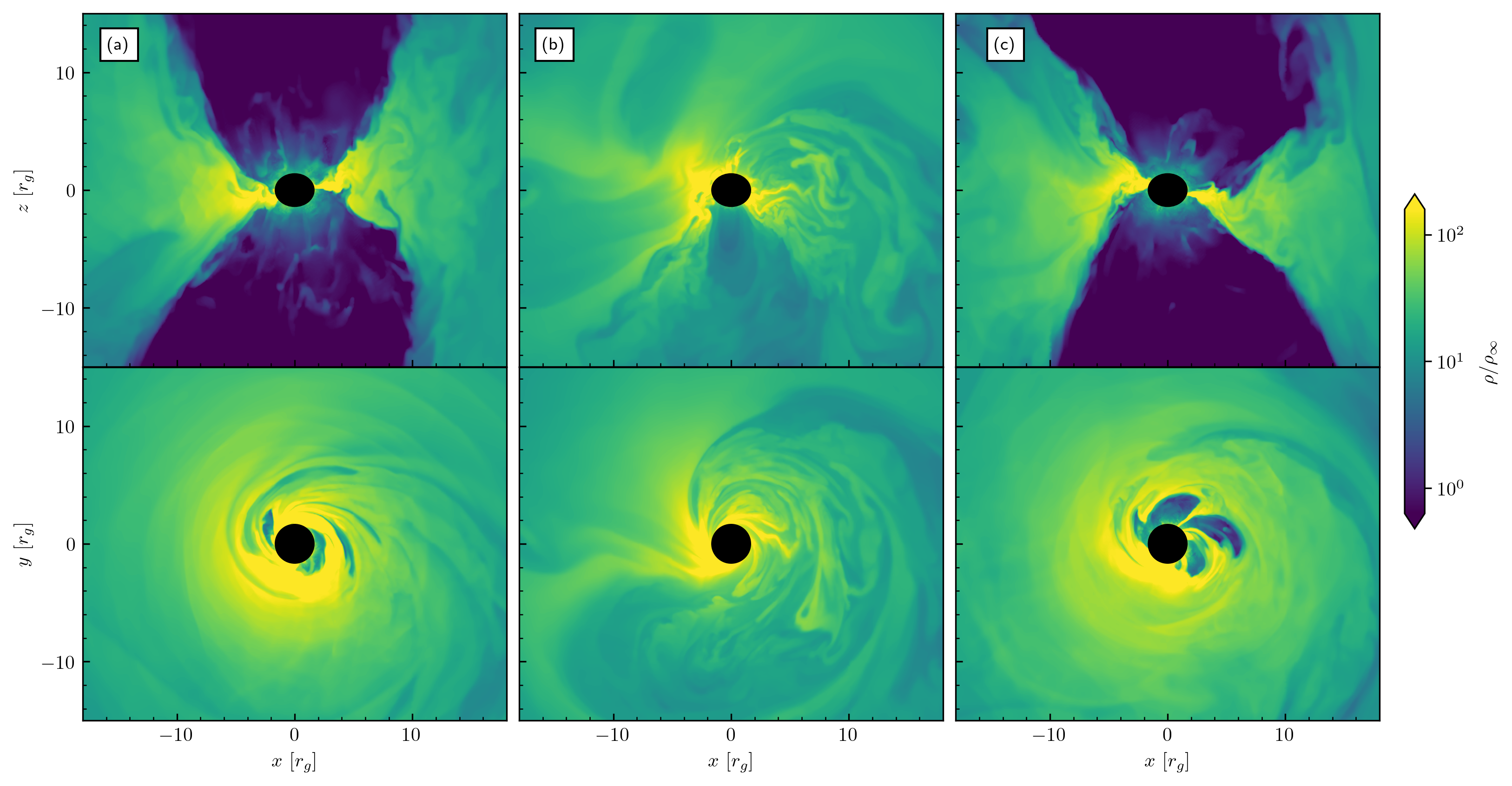}
\includegraphics[width=0.97\linewidth]{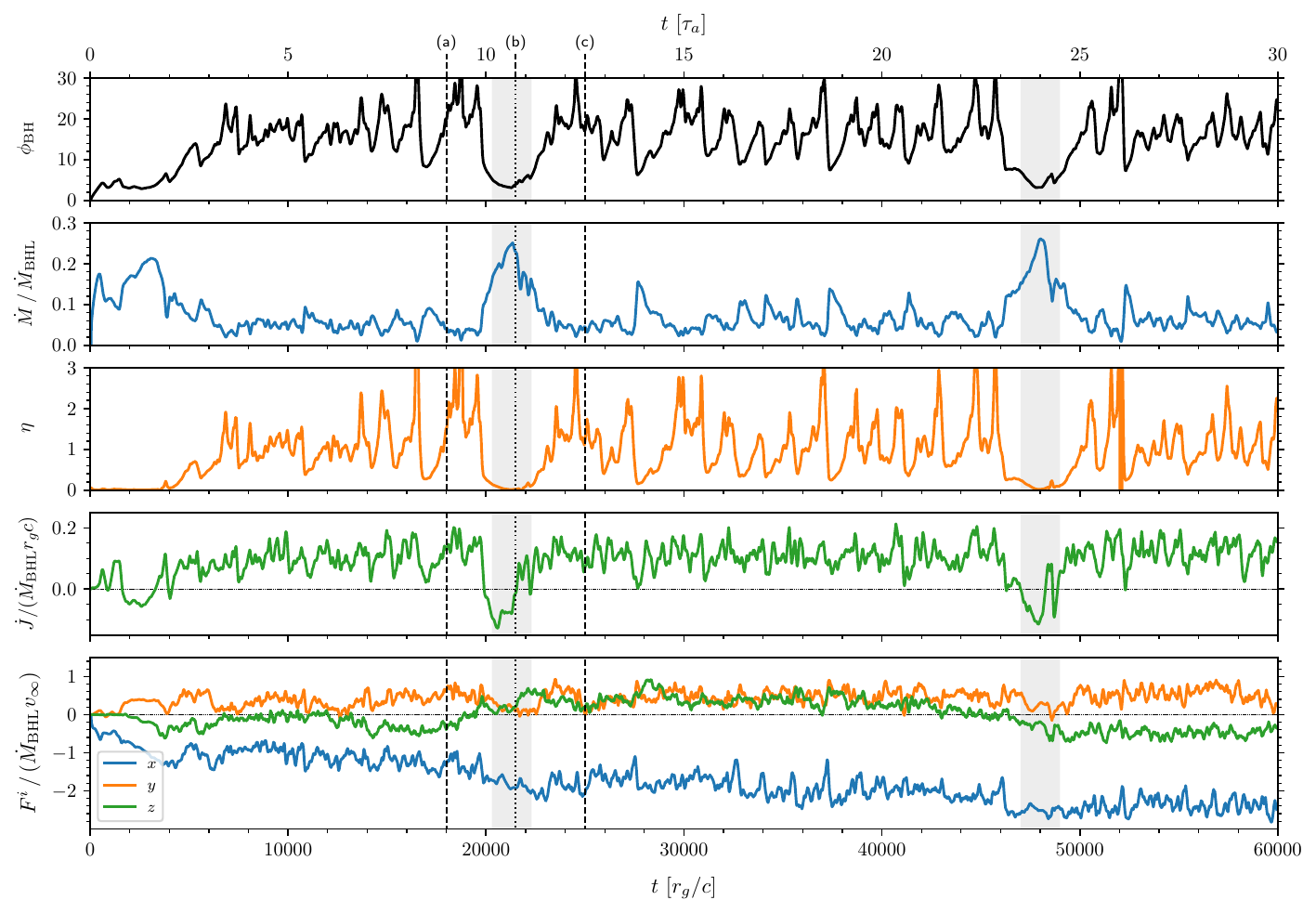}
\caption{Time evolution of physical quantities for the representative
    ($\beta_{10}$-$\theta_{90}$-$R_{200}$) model.
    Each panels of the line plots, from top to bottom, presents the
    dimensionless horizon magnetic flux $\phi_\text{BH}$, mass accretion rate
    $\dot{M}$, {energy outflow} efficiency $\eta$, angular momentum flux
    $\dot{J}$, and the total drag force $F^i$ over time. Quiescent periods
    (shaded) with a duration {$\sim 2000r_g/c$} are separated by an epoch of
    continued flux eruptions lasting {$\sim 24000r_g/c$}.
    Color plots on top of this figure show the mass density, $\rho$, on the $xz$
    (first row) and $xy$ (second row) plane in active states $t = 1.8\e{4}r_g/c$
    (a), $t =2.5\e{4}r_g/c$ (c) and a SANE-like quiescent state
    $t_=2.15\e{4}r_g/c$ (b), which are indicated with dashed and dotted vertical
    lines in the bottom panels. }
\label{fig:fiducial-timeseries}
\end{figure*}

\begin{figure*}
\centering
\includegraphics[width=\linewidth]{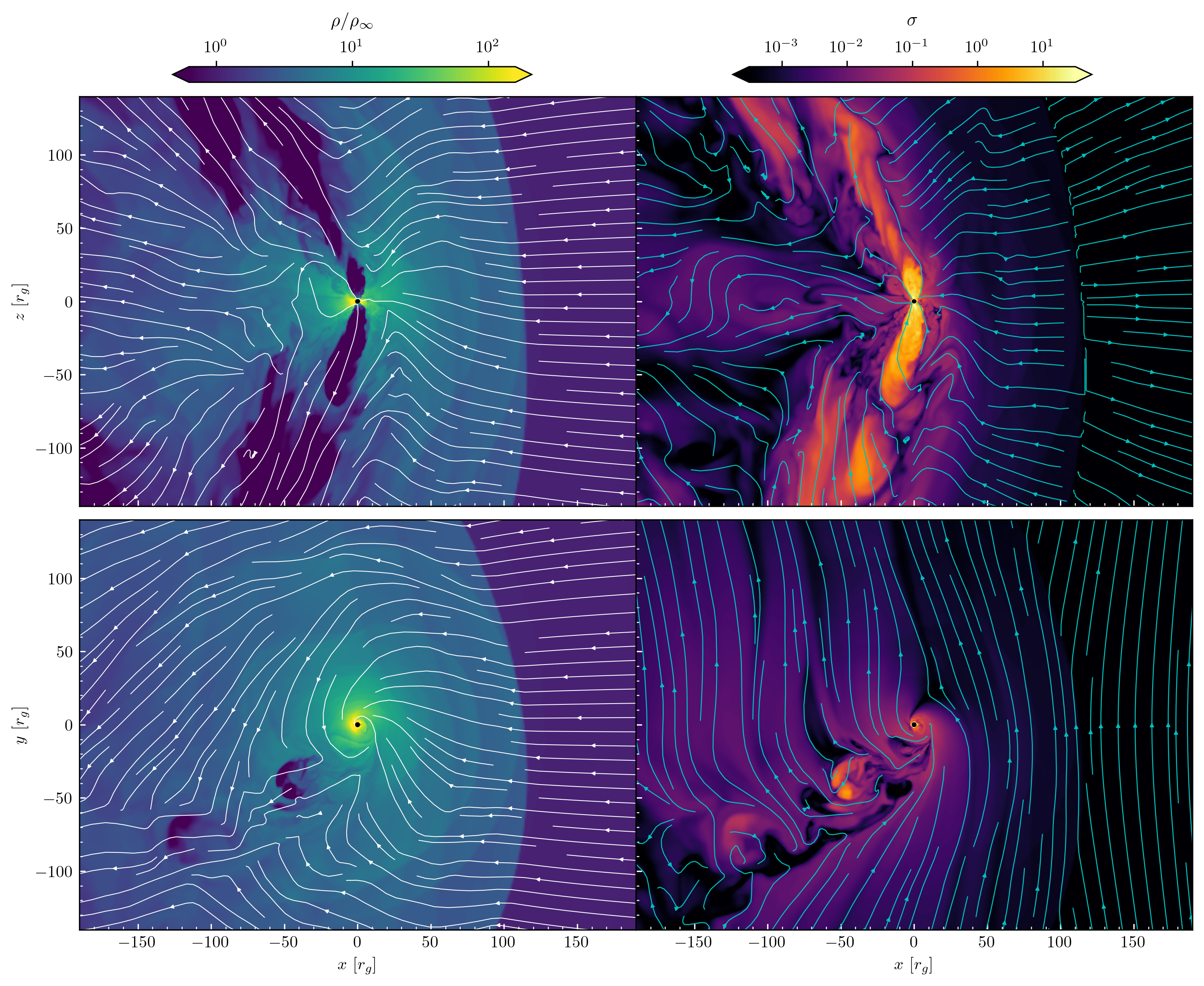}
\caption{Transient jets and magnetic flux eruptions in GRMHD
    Bondi-Hoyle-Lyttleton accretion. Shown here are mass density, $\rho$, (left)
    and relativistic magnetization, $\sigma$, (right) on the meridional plane
    (top) and the equatorial plane (bottom) from the fiducial model
    $\beta_{10}$-$\theta_{90}$-$R_{200}$ at $t=20,400 \, r_g/c$. The in-plane
    velocity and magnetic fields are shown with white and cyan streamlines on
    the left and right panels, respectively.}
\label{fig:fiducial-slice-big}
\end{figure*}

{\subsubsection{Initial jet launching process}}

Shortly following the beginning of the simulation, a shocked region around the
BH quickly expands and forms a bow-shaped shock front, which is a common
characteristic of supersonic BHL accretion flows
\cite[{e.g.,}][]{Foglizzo:2005in,Edgar:2004mk,Shima1985,Font:1998}. This
initial accretion phase is largely hydrodynamical. {Over time, the spin of
the BH drags the flow into rotation around the horizon, starting to form an
accretion disk, and magnetic flux is accumulated around the BH.}

{In Figure~\ref{fig:jet-buildup}, we show simulation snapshots at an early
time $(t \lesssim 2.5 \tau_a)$ focusing on the evolution of magnetic fields near
the BH. While the magnetic field is initially perpendicular to the BH spin,
turbulent fluid motion developing in the accretion flow generate vertical
(poloidal) magnetic flux, akin to an MHD dynamo. This can be seen from the third
and fourth rows of Fig.~\ref{fig:jet-buildup}, showing the vertical magnetic
field ($B^z$) on the equatorial plane. As the turbulent dynamo continues to
supply vertical magnetic fluxes into the accretion flow with various eddy sizes,
we find that a coarse large-scale magnetic flux separation emerges at $t\approx
1.8 \tau_a$ (the fifth column in Fig.~\ref{fig:jet-buildup}, appearing as a
two-sided spiral shape). This serves as a reservoir of unilateral vertical
magnetic field lines attached to the BH, which enables jet launching via the
Blandford-Znajek mechanism \cite{Blandford:1977ds}}.

{From the time series plots in Fig. \ref{fig:fiducial-timeseries}, we can
see that the magnetization around the BH quickly reaches half-MAD ($\phi_B\simeq
25$) levels within $t\approx 3 \tau_a$, as indicated by the dimensionless
magnetic flux $\phi_{\rm BH}$,} where a magnetically arrested disk (MAD) state
is characterized by $\phi_{\rm BH} \simeq 50$ \cite{Tchekhovskoy:2011zx}. This
built-up of magnetic flux coincides with the formation of a magnetized polar
funnel region near the black hole {as described above.}

\subsubsection{{Large-scale flow} morphology}

\reffig{fig:fiducial-slice-big} shows hydrodynamic and magnetic properties of
the accretion flow on the meridional ($xz$) and equatorial ($xy$) plane at
$t=20400r_g/c\approx 10 \tau_a$. Relativistic jets powered by the BH appear as
low-density, highly magnetized $(\sigma \gg 1)$ funnel regions traversing the
shock cone. The wind from the upstream collides with the jets and deflect them
downstream with its ram pressure \cite{Kaaz:2023b}. The amount of jet bending
correlates directly with the wind speed (see Fig. \ref{fig:speed} and the
accompanying discussion). {A pair of these bent, relativistic jets are
observed from tailed radio sources such as bent-tail radio galaxies
\cite[e.g.][]{Hardcastle:2003nm,1986ApJ...301..841O,Miley1975,Giacintucci:2009jx}.}

An accretion disk forms around the BH with the same direction of rotation as the
BH spin, and spans out to $r\sim 30 r_g$ on the equatorial plane although its
spatial extent is varying over time. At the outermost radius of the disk, its
circulatory flow is mixed with the regular downstream flow entering the shock
cone, forming a stagnation point, as can be seen from the {distribution of}
velocity streamlines {near $(x,y) = (-20, -30)r_g$ of the} lower left panel
of \reffig{fig:fiducial-slice-big}. {The matter inflow entering the shock
cone on the co-rotating side ($+y$) is smoothly connected to the circulatory
accretion flow, where that on the counter-rotating side ($-y$) collides with the
accretion flow to be pushed radially outward.} This leads to the bulk motion of
the downstream flow being deflected into $-\hat{y}$ direction. In contrast to
accretion simulations initialized or supplied with a finite angular momentum
({e.g.,} Fishbone-Moncrief torus \cite{Fishbone1976}, or BHL accretion
scenarios with a density/velocity gradient
\cite{Lora-Clavijo:2015,Cruz-Osorio:2016abh}), the incoming flow has zero net
angular momentum {in our setup}. Spin-induced frame dragging of the BH is
the only source of angular momentum imparted onto the flow, naturally limiting
the radial size of the accretion disk.

As pointed out in the discussion around Fig. \ref{fig:fiducial-timeseries}, the
BH is in a near MAD state \cite{Igumenshchev:2003rt,Narayan:2003by}. MAD states
feature the built-up of strong magnetic flux near the horizon, leading to the
establishment of a magnetically arrested flow structure near the innermost
stable circular orbit. Once the horizon magnetic flux rises above a threshold,
reconnection triggers an ejection of magnetic flux bundles from the BH
magnetosphere, accompanied by decay of the horizon magnetic flux
\cite{Ripperda:2020bpz,Ripperda:2021zpn,Chatterjee:2022mxg}. Shearing
instabilities, such as the Rayleigh-Taylor instability \cite{Kulkarni:2008vk},
ultimately trigger a magnetic flux eruption with a simultaneous mass accretion
inward, via interchanging magnetically buoyant low-density bubbles with less
magnetized, dense parcel of fluid \cite{Igumenshchev:2007bh}. Through this
process a MAD state is re-established over the viscous timescale of the
accretion flow. As a consequence, the BH does not sustain steady outflows but
exhibits fluctuations in the jet power and intermittent flux eruptions.

The quasi-periodic MAD cycle results in two notable features in the flow
morphology compared to unmagnetized axisymmetric BHL accretion flows.

\begin{figure*}
\centering
\includegraphics[width=0.95\linewidth]{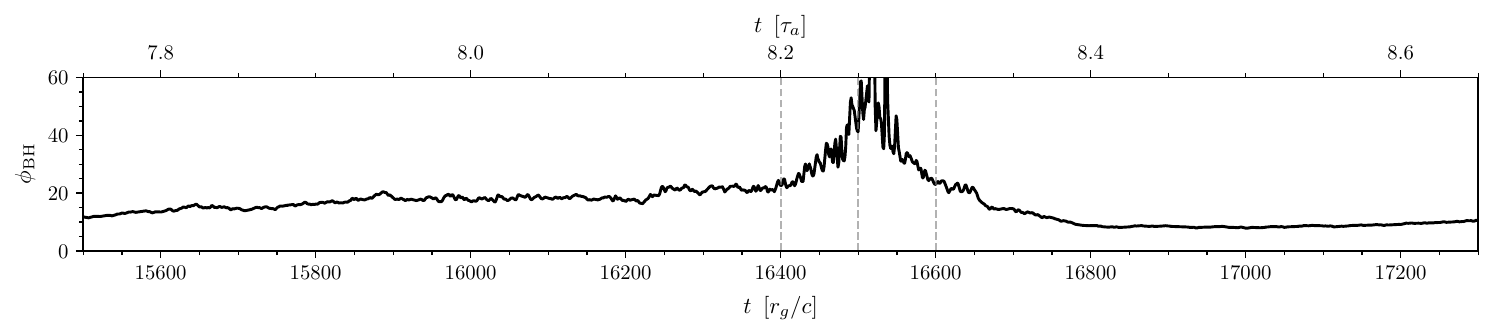}
\includegraphics[width=0.46\linewidth]{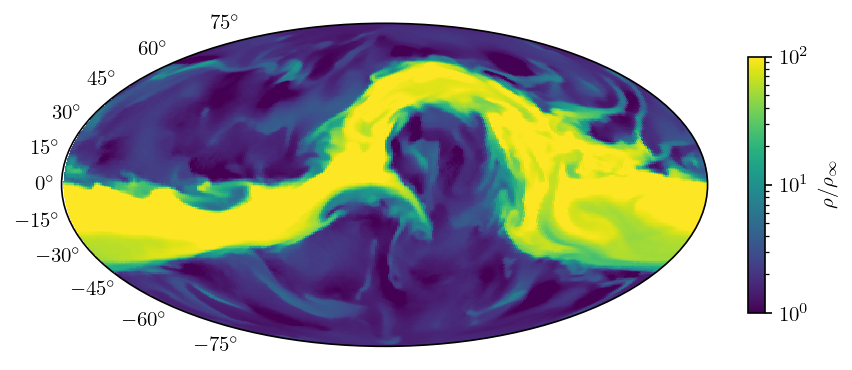}
\hspace{1ex}
\includegraphics[width=0.46\linewidth]{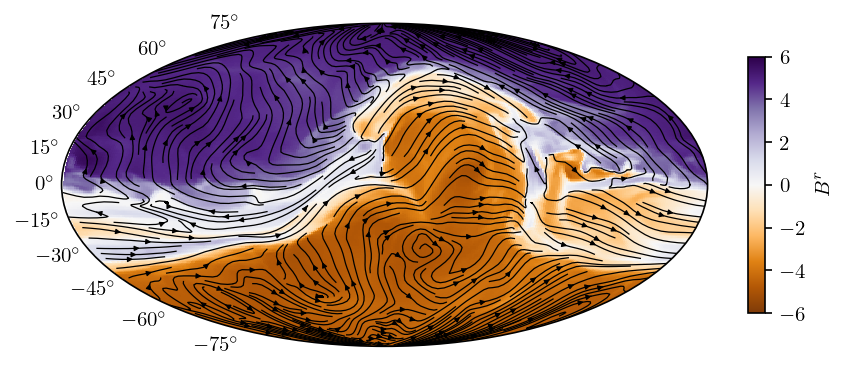}
\includegraphics[width=0.46\linewidth]{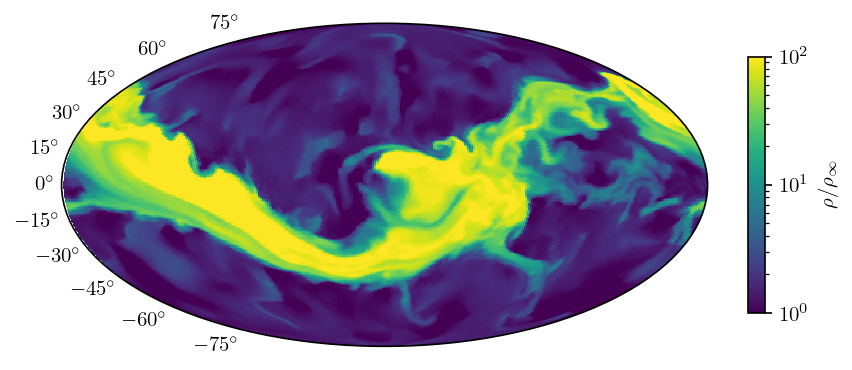}
\hspace{1ex}
\includegraphics[width=0.46\linewidth]{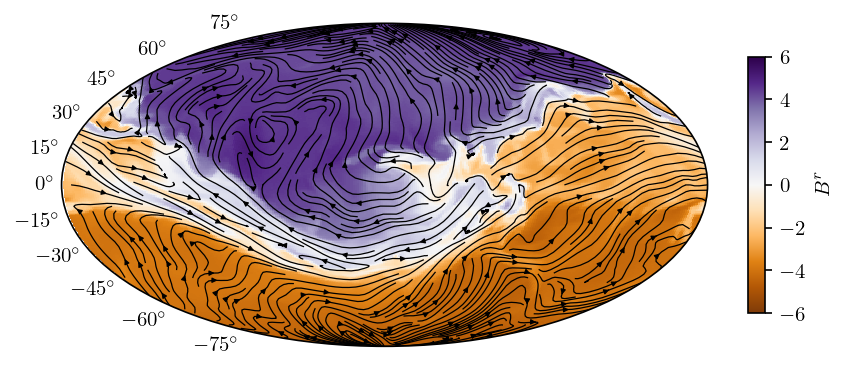}
\includegraphics[width=0.46\linewidth]{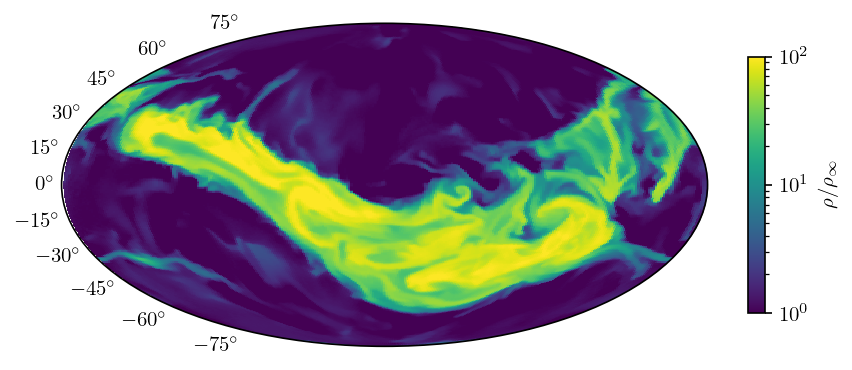}
\hspace{1ex}
\includegraphics[width=0.46\linewidth]{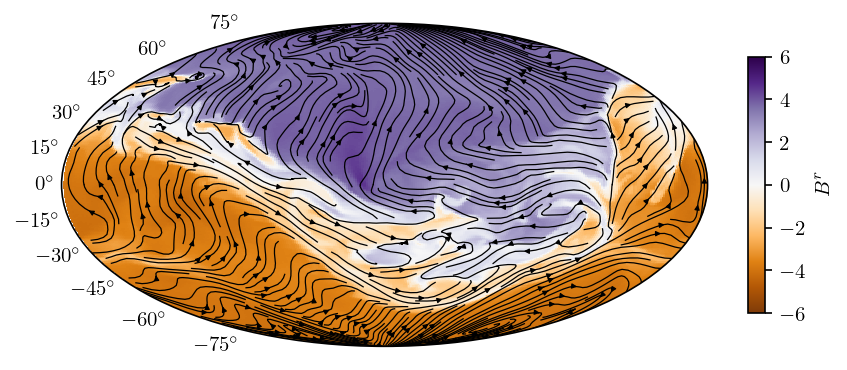} \\[2ex]
\includegraphics[width=0.32\linewidth]{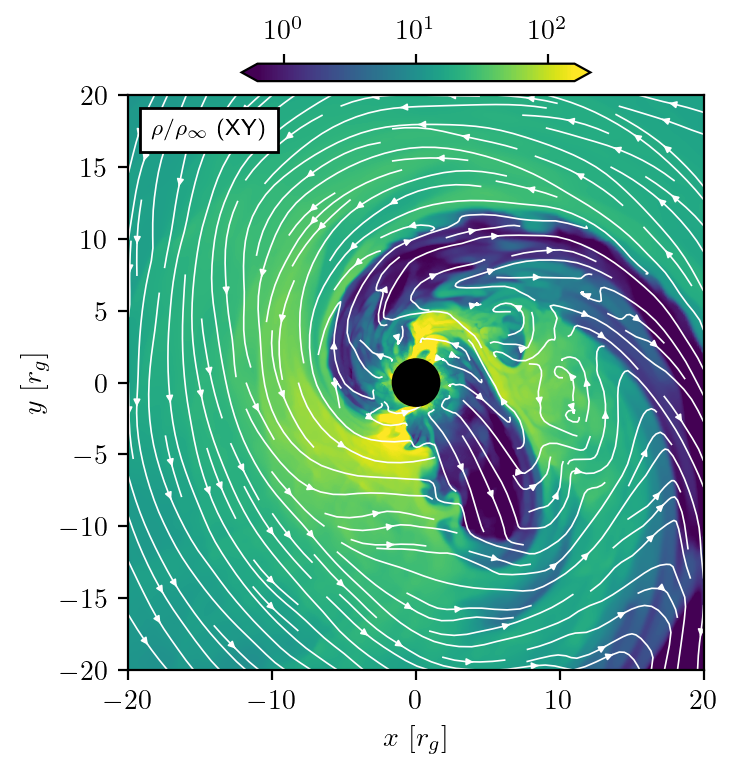}
\includegraphics[width=0.32\linewidth]{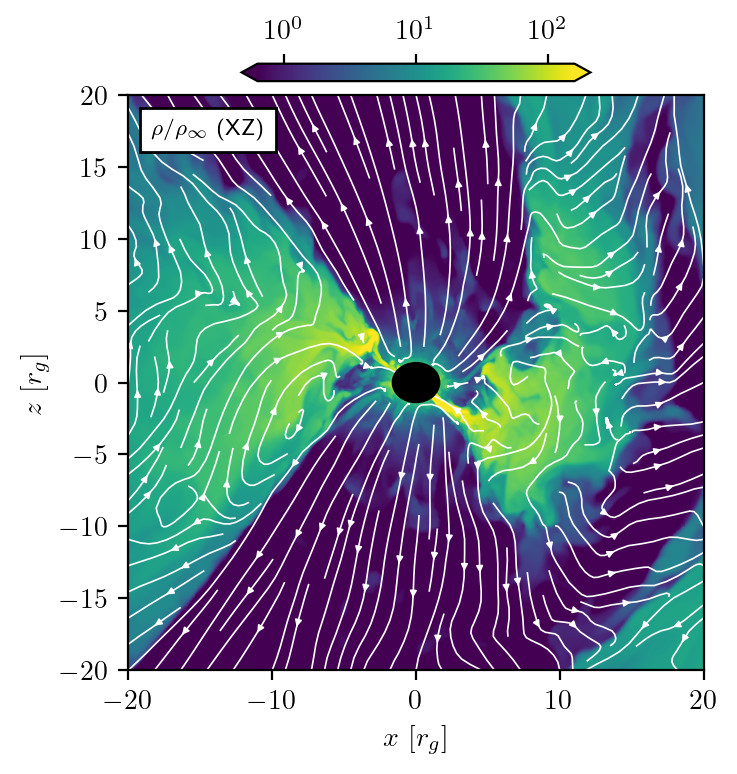}
\includegraphics[width=0.32\linewidth]{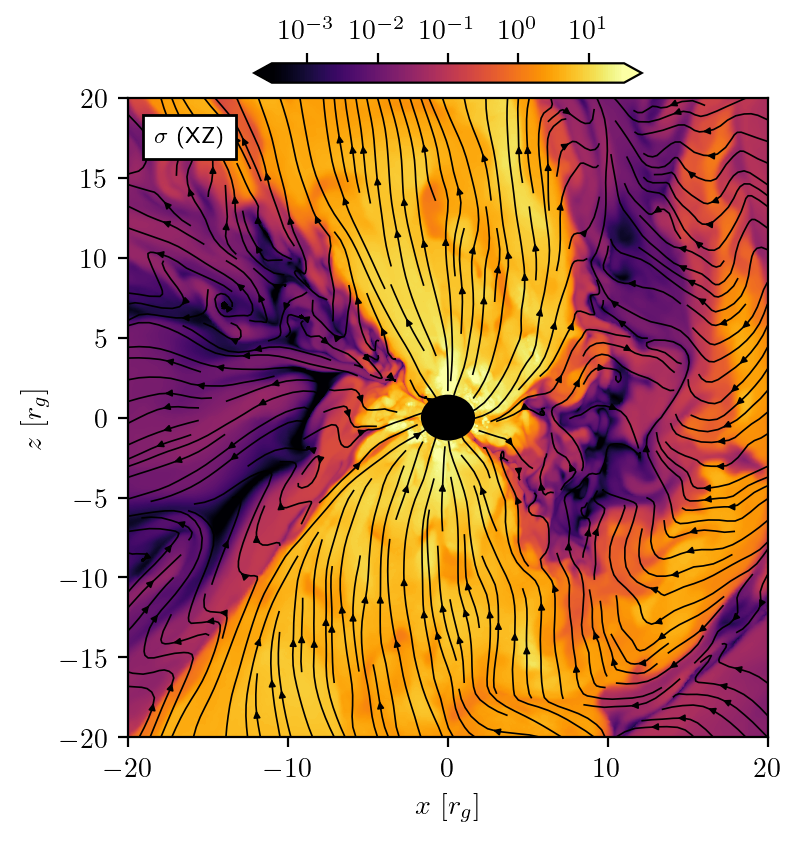}
\caption{A magnetic flux eruption event at $t=16500 r_g/c$.
    Top panel: change of dimensionless magnetic flux $\phi_\text{BH}$ on the
    black hole (BH) over the eruption. Three vertical dashed lines mark
    $t=16400, 16500, 16600 \, r_g/c$ respectively. Data points are displayed
    without smoothing.
    Middle panels: mass density, $\rho$, (left) and radial magnetic field $B^r$
    {in units ($[\rho_\infty^{1/2} c]$)}, (right) on a spherical
    surface $r=5r_g$ are shown with the Mollweide projection aligned with the BH
    spin axis. The radial component of the magnetic field is shown in color and
    the angular components are shown with black streamlines. The center of the
    plot corresponds to the $+\hat{x}$ direction. From top to bottom, each row
    corresponds to $t=16400, 16500, 16600 \, r_g/c$.
    Bottom panels: mass density on the equatorial plane (left), on the
    meridional plane (center), and the relativistic magnetization $\sigma$ on
    the meridional plane (right) at $t=16500 r_g/c$. In-plane velocity and
    magnetic field are shown with white and black streamlines.}
\label{fig:eruption-event}
\end{figure*}

\begin{enumerate}[label=(\arabic*)]
\item Each eruption event launches a pressure wave that expands outward from the
    black hole. This wavefront with a relatively higher density pushes the
    boundary of the bow shock further upstream, expanding the shock cone, before
    it shrinks back due to the ram pressure of the incoming fluid. As a
    consequence, the bow shock exhibits a breathing motion in lockstep with each
    of the magnetic flux eruptions. This results in a deformed morphology of the
    shockfront, rather than a smooth parabolic shape commonly observed from
    unmagnetized cases
    \cite[{e.g.,}][]{Shima1985,Penner:2013,Lora-Clavijo:2013,Gracia-Linares:2015woa}.
    This eruption-driven expansion of the bow shock and multiple pressure waves
    approaching the bow shock can be seen from the left panels of
    \reffig{fig:fiducial-slice-big} or in \reffig{fig:jet-3d}.
\item A series of magnetic flux tubes with high magnetization are formed near
    the BH and drift downstream on the equatorial plane as they are released, a
    few of which can be identified from the region $x<0$, $y<0$ in the lower
    panels of \reffig{fig:fiducial-slice-big}. This situation is reminiscent of
    unboosted MAD BH flows
    \cite[{e.g.,}][]{Chatterjee:2022mxg,Ripperda:2021zpn,Porth:2020txf}.
    When a flux eruption event occurs, magnetic pressure pushes the matter
    outward via an interchange instability and forms highly magnetized, hot,
    low-density voids near the horizon ({e.g.,} see the simulation snapshots
    included in \reffig{fig:fiducial-timeseries} and
    \reffig{fig:eruption-event}). The ejected flux tubes are spiralling outward
    from the BH \cite{Porth:2020txf}, getting sheared and ultimately reach the
    stagnation point, where they are fragmented by ram pressure and released
    into the downstream flow as a mushroom-shaped feature. In magnetic flux
    eruptions of axially symmetric accretion flows, flux tubes would have been
    launched without a preferred direction. Here, due to a preferential
    direction of the ambient flow, the flux tubes can be only released toward a
    particular range of angle relative to the direction of the upstream wind.
    The ejected flux tubes are filled with hot non-thermal plasma produced by
    the near-horizon magnetic reconnection that initially created the tubes
    \cite{Ripperda:2020bpz,Ripperda:2021zpn}. This plasma has the potential to
    power TeV and X-ray flares \cite{Porth:2020txf,Hakobyan:2022alv}, in a
    similar manner to what has been proposed as an explanation for galactic
    flares near Sgr A* \cite{Dexter:2020cuv,Antonopoulou:2025tgs}.
\end{enumerate}

\subsubsection{Time evolution} \label{sec:time-evolution}

Having described the overall properties of the accretion flow, we now focus on
its time variability. In \reffig{fig:fiducial-timeseries}, we provide the time
series data of the dimensionless horizon magnetic flux $\phi_\text{BH}$, mass
accretion rate $\dot{M}$, {energy outflow} efficiency $\eta =
(\dot{M}-\dot{E}/c^2) / \dot{M}$, (outward directed) angular momentum flux
$\dot{J}$, and the total drag force $F^i$. Based on these, we examine the time
evolution of our fiducial accretion flow setup in this section.

After the initial development of relativistic polar outflows from the BH ($t\sim
3 \tau_a$), the accretion flow undergoes an oscillatory MAD cycle showing a
transient behavior of $\phi_{\rm BH}$ associated with episodes of magnetic flux
eruptions. In active phases, when the jet is present, the mass accretion rate is
suppressed, whereas it is enhanced in quiet (accreting) phases when the jet is
absent. The overall evolution observed in this model is broadly consistent with
previous studies on MADs
\cite[e.g.][]{White:2019csp,Porth:2020txf,Chatterjee:2022mxg,Tchekhovskoy:2011zx}.
The continued eruption cycle is maintained until $t\sim 10\tau_a$, which we
hereafter refer to as the first eruption epoch.

Following the final flux eruption of the first eruption epoch at $t \sim 10
\tau_a$, the magnetic flux of the BH drops and the system enters a fully
quiescent period, $10 \tau_a \leq t \leq 12 \tau_a$ with the jet being fully
quenched. The accretion flow temporarily enters a standard and normal evolution
\cite[SANE,][]{Narayan:2012yp} regime {in which magnetized polar funnel are
replaced by a smooth inflow of matter toward the BH and the accretion exhibits a
relavitely laminar flow on the equatorial plane (with only mild turbulent
eddies) without a strong vertical stratification \cite{Chatterjee:2022mxg}.} As
a result, the mass accretion rate is notably increased and the BH experiences a
temporary spin up ($\dot{J} < 0$) from the falling matter.

In the middle of the SANE-like quiescent period, the horizon magnetic flux and
the {energy outflow} efficiency rises again, mass accretion decreases, and
the angular momentum flux transitions from inflow ($-$) to outflow ($+$). The
evolution of the system in this time window is similar to the process of the
first jet launching phase, indicating the revival of it. This revival process
takes roughly a viscous time scale of the disk. The system re-enters the MAD
state and the jet is launched again at $t\sim 12 \tau_a$, which marks the
beginning of the second eruption epoch lasting $12\tau_a \leq t \leq 23 \tau_a$.

In the upper half of \reffig{fig:fiducial-timeseries}, we include simulation
snapshots displaying the mass density distribution of the accretion flow at
three different times $t_{(a)}=1.8\e{4}r_g/c$, $t_{(b)}=2.15\e{4}r_g/c$, and
$t_{(c)}=2.5\e{4}r_g/c$, each of which corresponds to $(a)$ a {MAD-like}
active state by the end of the first eruption epoch, $(b)$ a quiet
{SANE-like} state with no jet present, and $(c)$ an active state in the
second eruption epoch after the revival. This process overall repeats
periodically.

\subsection{Magnetic flux eruptions}
\label{sec:eruptions}

In the preceding sections, we have described the global dynamics leading to the
establishment of a MAD accretion state and subsequent magnetic flux eruption
events. In the following, we provide a detailed description of the flux eruption
event at $t=16500 r_g/c$ as a representative example of the process, and
elucidate its effect on the jet morphology.

\subsubsection{Near black-hole dynamics}

Along with the three sections of plots comprising \reffig{fig:eruption-event},
we now illustrate a comprehensive picture of a single magnetic flux eruption event.

\begin{itemize}[leftmargin=3ex, labelsep=1.5ex]
\item \emph{Evolution of the horizon magnetic flux}
    (\reffig{fig:eruption-event}, top panel): The eruption cycle begins with a
    magnetically relaxed state after the previous eruption has settled down. The
    horizon magnetic flux $\phi_\text{BH}$ starts to increase from $t=15500
    r_g/c$, reaching $\phi_\text{BH}\approx 20$ at $16300 r_g/c$. Then it
    rapidly rises to $\phi_\text{BH} > 50$ around $t=16500 r_g/c$, and relaxes
    down to $\phi_\text{BH}\approx 10$ at $t=16700 r_g/c$ after the eruption. 
\item \emph{Nutation of the accreting plane} (middle panels in the Mollweide
    projection in \reffig{fig:eruption-event}{):} These panels, showing mass
    density and magnetic field on a spherical surface $r=5r_g$ at
    $t=16400r_g/c$, $16500r_g/c$ and $16600r_g/c$, visualize angular
    distribution of the accretion flow and the geometry of magnetic field near
    the BH during the eruption event. In the pre-eruption stage, the accretion
    disk develops precession with an increasingly large amplitude over time.  
    Left column of the panels shows that the accretion disk is subject to a
    significant tilt and distortion during the eruption (which otherwise would
    appear as a smooth strip along the equator). The ejection of the magnetic
    flux tube during the eruption is off the equator and exerts a strong recoil
    (and corresponding torque) onto the system, which leads to a nutation of the
    accretion disk.
\item \emph{Tearing of accretion disk} (bottom panels of
    \reffig{fig:eruption-event}{):} The asymmetric ejection of the magnetic
    flux tube (low density/high magnetization region) and the resulting recoil
    strongly tilting and tearing the inner accretion flow can be clearly seen
    from the figure. The dynamical time scale of the fluid at which the
    accretion disk is torn can be estimated via the local Keplerian orbital
    period as
    \begin{equation}
        T \approx \frac{2\pi}{\Omega_K} = 2\pi \left(\frac{r^3}{GM}\right)^{1/2}
            = 370 r_g/c
    \end{equation}
    for $r=15 r_g$, which is in good agreement with the disk precession period
    $\approx 400 r_g/c$, which we estimate {empirically.}
    While the inner part of the accretion disk has a shorter dynamical timescale
    ($r < 15 r_g$) than the driving frequency ($T \approx 400 r_g/c$) and is
    able to reorganize itself, the outer part of the disk with a longer
    dynamical timescale ($r > 15r_g$) cannot dynamically react (synchronize) to
    the driving and is decoupled from the inner part. In a different context,
    highly tilted accretion disks have been observed to suffer more tearing,
    precession, and fragmentation, when the inner part of the disk is subject to
    torque \cite{Liska:2019vne}.
\end{itemize}

\subsubsection{Jet morphology}

\begin{figure*}
\centering
\subfloat[$t=16500r_g/c$]{\includegraphics[width=0.49\linewidth]{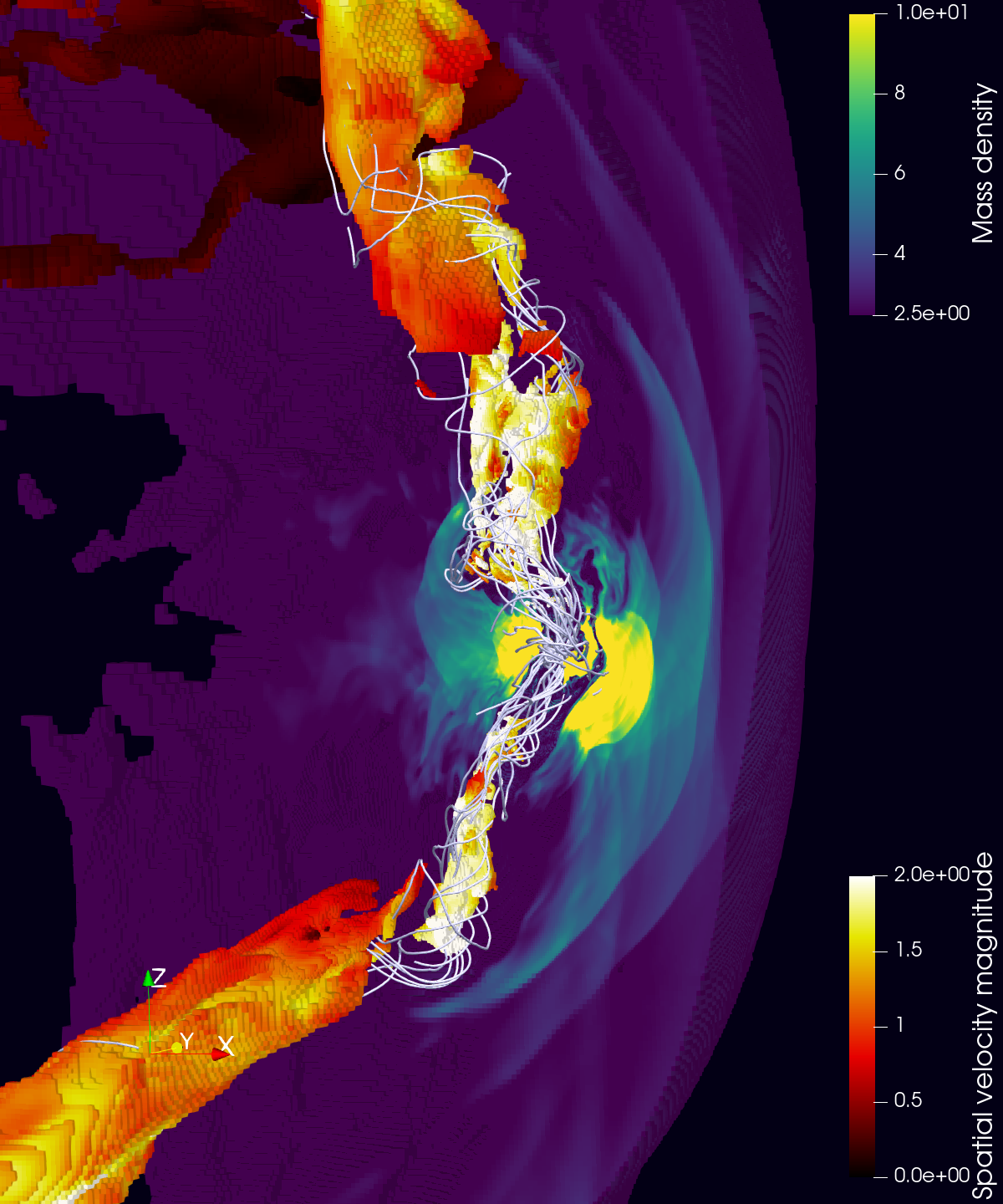}}
\hfill
\subfloat[$t=21500r_g/c$]{\includegraphics[width=0.49\linewidth]{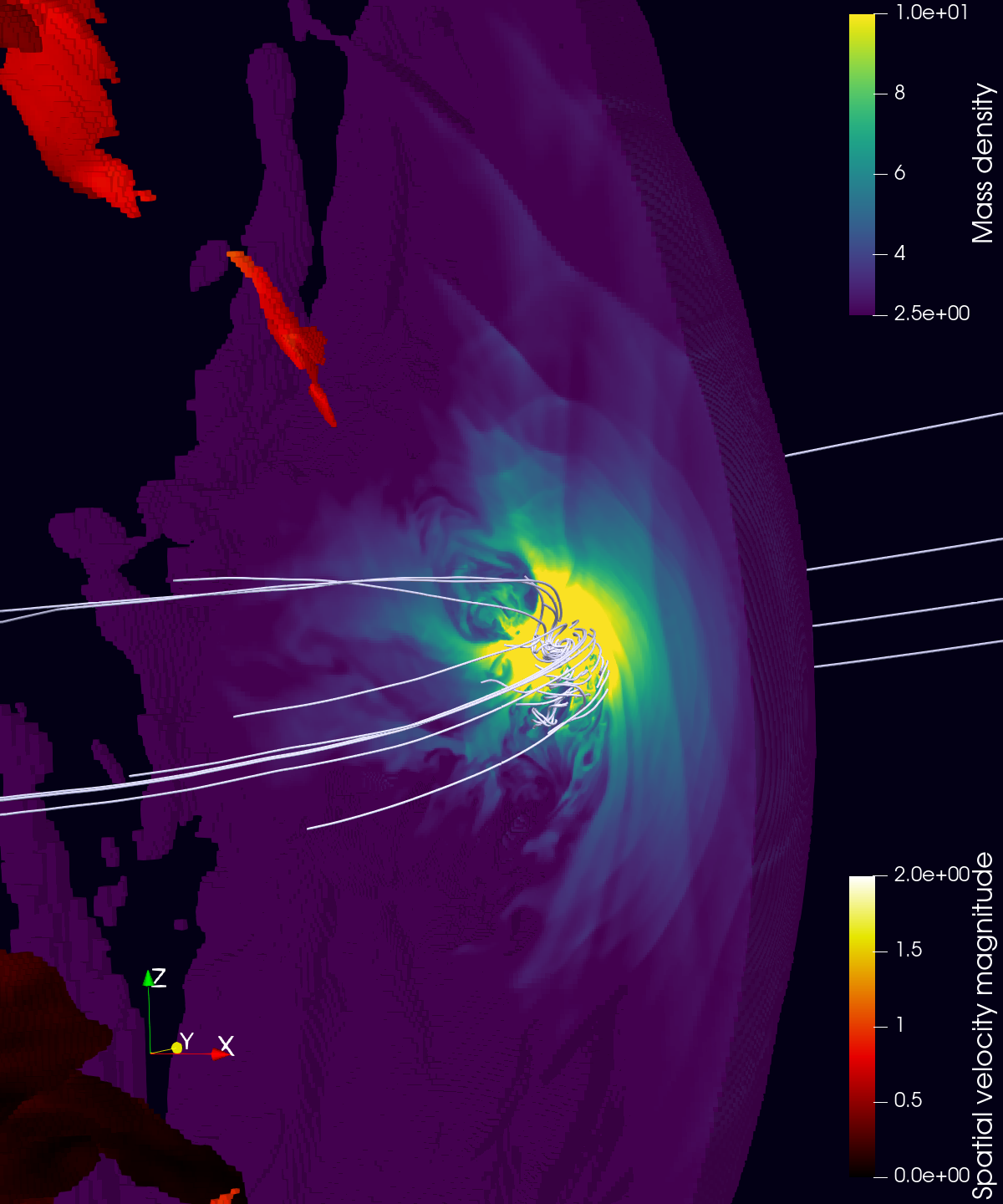}}
\caption{A three-dimensional rendering of the simulation depicting an active
    state (left) and a quenched state (right). Shown here are low-density,
    highly magnetized outflow (filtered by $\rho < 0.05\rho_\infty$, red-white
    colors), accretion flow inside the bow shock (filtered by $\rho > 2
    \rho_\infty$ and half-cut to vertical, blue-yellow colors), and magnetic
    field lines emanating from the accretion disk. Each colormap shows the
    magnitude of the spatial components of the four-velocity and the normalized
    mass density $\rho/\rho_\infty$. A vertical tearing of the accretion disk
    (see \refsec{sec:eruptions}) is visible in the left panel.}
\label{fig:jet-3d}
\end{figure*}

\begin{figure*}
\centering
\includegraphics[width=0.33\linewidth]{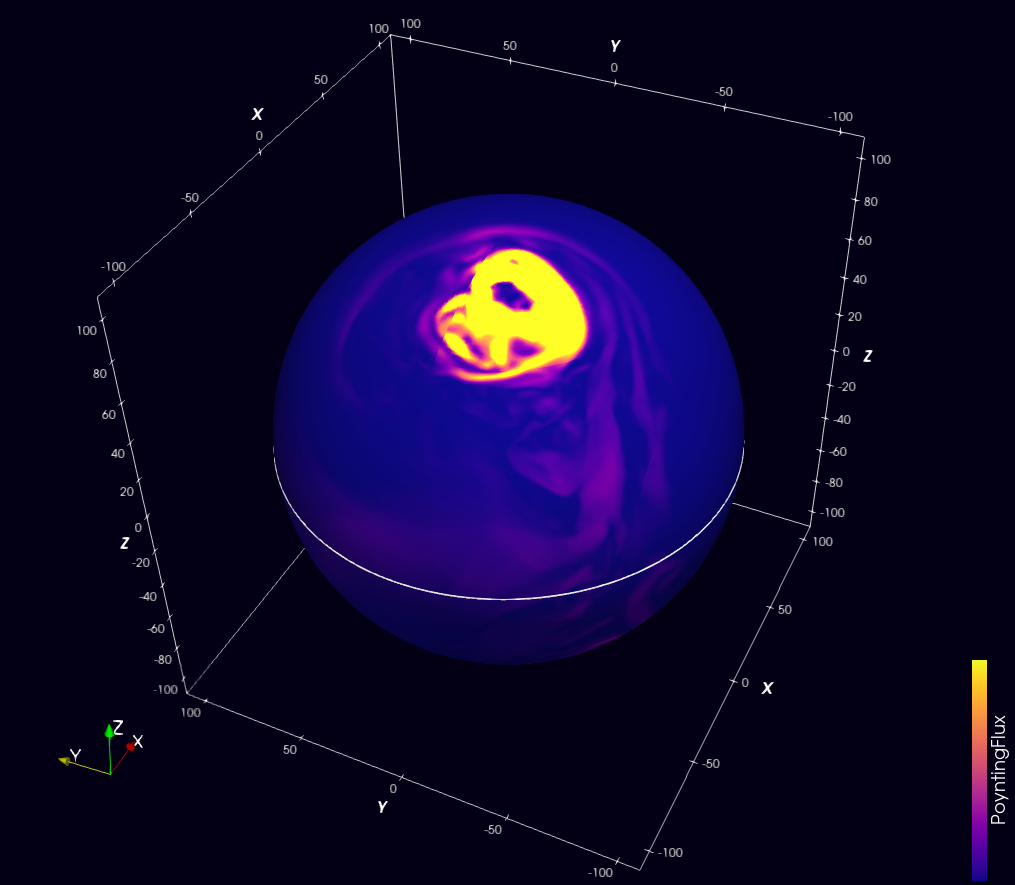}
\hspace{-1ex}
\includegraphics[width=0.33\linewidth]{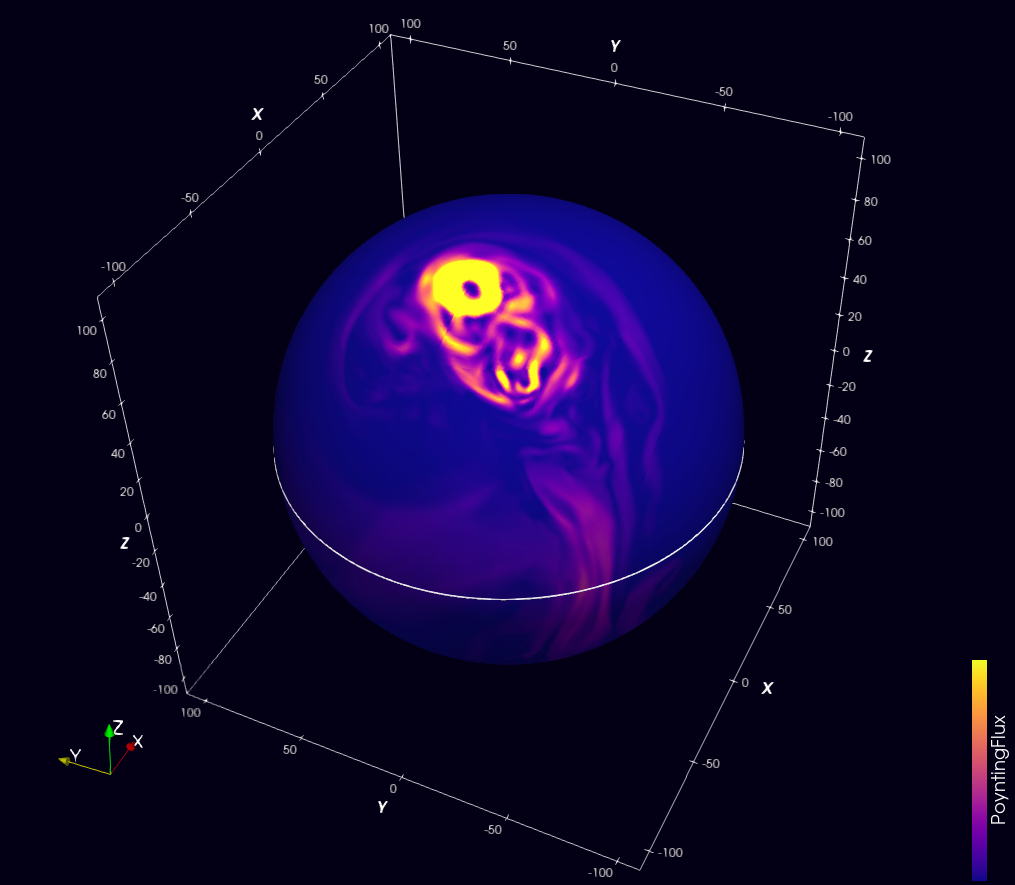}
\hspace{-1ex}
\includegraphics[width=0.33\linewidth]{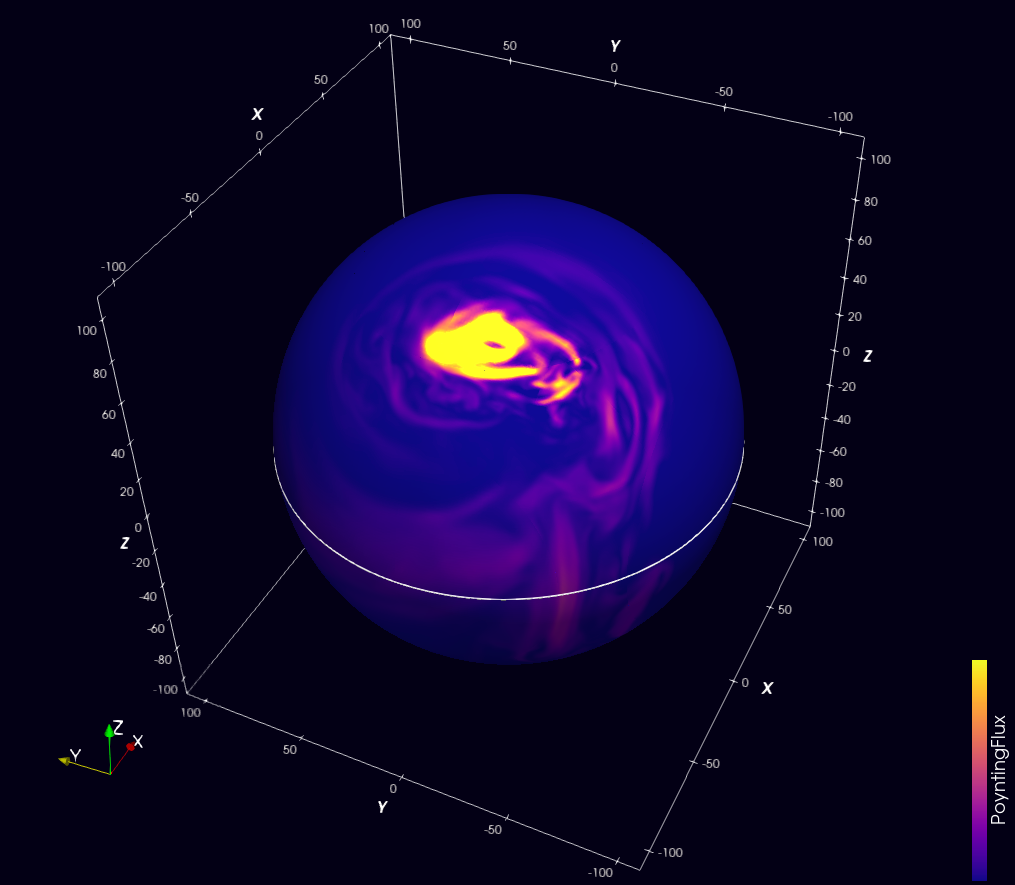}
\caption{Distribution of the radial Poynting flux, $\left(T_{\rm
    EM}\right)^t_r$, on the upper hemisphere of a spherical surface $r=100r_g$
    at $t=19600\, r_g/c$ (left), $t=19800\, r_g/c$ (center), and $t=20000\,
    r_g/c$ (right).}
\label{fig:poynting-3d}
\end{figure*}

High-resolution simulations of tilted accretion disks have shown that the
motions of the disk and the jet are coupled \cite{Liska:2017alm,Liska:2019vne},
{and we observe a similar phenomenon in our simulations. While the structure
of the jet remains more or less aligned with the BH spin very close the horizon
($r\lesssim 5 r_g$),} the rapid nutation of the accretion disk {and a
consequent wobbling of the the polar funnels surrounded by the accretion flow}
results in fluctuations in the direction of the jet {at $r\gtrsim 10 r_g$
(see also Ref. \cite{Liska:2017alm})}.
From a three-dimensional visualization in \reffig{fig:jet-3d}, one can observe
the twisted morphology of the jet associated with the nutation of the
{near-BH accretion flow}. {This non-smooth jet morphology may further
enhance kink-like instabilities naturally present in these systems
\cite{Appl2000,Li:2000mc,Narayan:2009cz,Bromberg:2015wra}.} At larger distances,
the interaction with the ambient wind gradually smooths out these features. We
can also see how the jet is quenched at a later time.

In \reffig{fig:poynting-3d}, we show the distribution of the radially outgoing
electromagnetic flux $\left(T_{\rm EM}\right)^t_r$ at $r=100r_g$. The opening
angle of the jet shows variations between $15^\circ$ and $25^\circ$, where the
position of the peak Poynting flux (center of the jet) precesses with an
amplitude $\lesssim 15^\circ$. While the jet still exhibits an oscillatory
behavior at this radius to some extent, its angular variations remain much
smaller than the inner accretion disk attached to the BH. We expect that these
variations will be further attenuated at a larger radius.

\subsection{Drag and spin-down}
\label{sec:drag}

Here we look into the transport of linear and angular momentum between the BH
and the accretion flow, which are responsible for the deceleration and spin-down
of the BH. {A time scale which we will frequently refer to for the
discussions in this section is the mass doubling timescale $\tau_M = M/\dot{M}$,
which can be rescaled using the BHL mass accretion rate $\dot{M}_{\rm BHL}$ as
\begin{equation} \label{eq:mass doubling timescale}
\begin{split}
    \tau_M & = 2.8\e{4}
    \left(\frac{\dot{M}}{\dot{M}_{\rm BHL}}\right)^{-1}
    \left(\frac{v_\infty}{1000 \, \mathrm{km \, s^{-1}}}\right)^3 \\
    & \times
    \left(\frac{\rho_\infty}{10^{-10} \, \mathrm{g \, cm^{-3}}}\right)^{-1}
    \left(\frac{M}{100 M_\odot}\right)^{-1} \text{ yr} \, .
\end{split}
\end{equation}
The normalized mass accretion rate is $\dot{M} / \dot{M}_{\rm BHL} \sim 0.1$ in
our simulation (see Fig.~\ref{fig:fiducial-timeseries}).}

\subsubsection{Drag force}

An accretor traveling through a gaseous medium is subject to a dynamical
friction \cite{1943ApJ....97..255C}. The reference scale of this drag is
{$\dot{M}_\mathrm{BHL} v_\infty$, which} is the drag force in the ballistic
(dust) limit, and a multiplicative correction factor needs to be included in
generic cases.\footnote{Newtonian studies suggest that the correction is not
expected to exceed a factor of ten in hydrodynamic BHL accretion
\cite{Edgar:2004mk}.} For convenience, we define the fudge factor
\begin{equation}
    f_\text{BHL}^i \equiv F^i / (\dot{M}_\mathrm{BHL} v_\infty)
\end{equation}
where $F^i$ is the measured drag force in each directions. Since our simulation
is performed in a fixed spacetime and does not consistently capture the
slow-down of the BH in dynamical general relativity, the drag force is estimated
by means of an approximate formula Eq.~\eqref{eq:total drag force}.

The bottom panel of \reffig{fig:fiducial-timeseries} shows the time variation of
the total drag force normalized with the BHL drag scale {$\dot{M}_{\rm BHL}
v_\infty$}, effectively displaying the factor $f^i_\text{BHL}$ for each spatial
directions. The tangential drag (dynamical friction) reaches a steady value of
$f_\text{BHL}^x \sim 2.5$ around $t =  5\e{4} r_g/c$. The linear momentum
accretion rate $F_\text{mom}^x$ is nearly zero when time averaged, yet it is
very oscillatory. It is the gravitational drag $F_\text{grav}^x$ that
constitutes a dominant net portion of the total drag. See
\refsec{sec:r-dependence} for the raw time series data of $F^x_\text{mom}$ and
$F^x_\text{grav}$. A similar trend is observed for $F^y$ and $F^z$ as well,
suggesting that the accretion flow plunging into the BH is mostly symmetric when
averaged over time. However, we note that $F^y_\text{mom}$ shows a small
positive average $\sim 0.1 \dot{M}_\mathrm{BHL} v_\infty$.

The drag force results in a slow-down of the BH relative to the surrounding
medium. The deceleration time scale can be estimated by dividing the initial BH
linear momentum by the drag force
{
\begin{equation} \label{eq:deceleration timescale}
    \tau_\mathrm{dec}
        = \frac{M v_\infty}{f_\text{BHL} (\dot{M}_\text{BHL} v_\infty)}
        = \left(f_\text{BHL} \right)^{-1} \tau_M
\end{equation}
where the mass doubling timescale $\tau_M$ is given as Eq.~\eqref{eq:mass
doubling timescale}. Plugging in $\dot{M} / \dot{M}_{\rm BHL} \sim 0.1$ and
$f_\text{BHL} \sim 2.5$ from our fiducial simulation, we get $\tau_{\rm dec}
\sim 10^5 \, {\rm yr}$ for the representative values of $v_\infty$,
$\rho_\infty$, $M$ chosen in \eqref{eq:mass doubling timescale}.}

On top of the drag force $F^x$ tangential to the direction of BH motion, there
exist nonzero transverse drag ($F^y$) and vertical drag ($F^z$) which are
perpendicular and parallel to the BH spin. These drags can potentially induce a
deflection or oscillatory features in the trajectory of the accreting BH. The
transverse drag force $F^y$ shows an average magnitude $\approx 0.5
\dot{M}_\text{BHL} v_\infty$, which is about 20\% of the tangential drag $F^x$.
The vertical drag $F^z$ has a similar magnitude as the transverse part $F^y$.
The transverse drag $F^y$ maintains a positive finite value where the vertical
drag periodically flips its sign between eruption epochs. Each of these
components are intimately related with the gravitational Magnus effect
(\refsec{sec:magnus}) and the magnetic reversal of jets (\refsec{sec:reversal})
observed in our simulations.

\subsubsection{Spin-down}

As previously noted, in our setup, the spin effect from the BH is the only
physical origin of circulation introduced in the accretion flow. From
\reffig{fig:fiducial-timeseries}, we see that the angular momentum transport
rate from the BH into accretion flows is $\dot{J} \approx 0.1 \dot{M}_\text{BHL}
r_g c$ during eruption epochs where it is reversed during quiescent periods.
Angular momentum in the disk is mainly transported in flux eruption episodes
\cite{Chatterjee:2022mxg}, where the overall spin-down of the BH is largely
affected by the jet as well \cite{Lowell:2023kyu}. {Since our problem setup
(BHL accretion) is different from a commonly referenced MAD configuration which
is being reached starting from an accreting torus, it is noted that an
additional feedback mechanism might contribute to the angular momentum
transport.}

{A systematic investigation on the spin-down of a BH in the MAD state showed
that the characteristic decaying time of the BH dimensionless spin {$Jc/GM^2 =
a/r_g$} is about $10\%$ of the mass doubling time scale \cite{Lowell:2023kyu}.
To explore the same aspect but for the BHL accretion, we estimate the spin-down
rate of the BH during an active phase in our simulation as follows. The angular
momentum loss timescale of the BH can be estimated as
\begin{equation} \label{eq:spin-down timescale}
    \tau_J = \frac{J}{\dot{J}}
     = \left(\frac{\dot{J}}{\dot{M}_{\rm BHL} r_g c}\right)^{-1}
     \left(\frac{\dot{M}}{\dot{M}_{\rm BHL}}\right)
    \left(\frac{a}{r_g}\right) \, \tau_M
     \,. 
\end{equation}
Our simulation results (Fig.~\ref{fig:fiducial-timeseries}) shows $\dot{J} /
(\dot{M}_{\rm BHL} r_g c) \approx 0.1$ and $\dot{M}/\dot{M}_{\rm BHL} \approx
0.05$, indicating $\tau_J \sim \tau_M/2$. Then the spin-down timescale of the
dimensionless spin $a/r_g$ can be estimated as
\begin{equation}
    \frac{a}{r_g} \propto \frac{J}{M^2} \propto \frac{e^{-t/\tau_J}}{e^{2t/\tau_M}}
    \sim e^{- 4 t / \tau_M} \, ,
\end{equation}
yielding a slightly longer spin-down timescale $\tau_M / 4$ than $\approx \tau_M
/ 10$ from Ref. \cite{Lowell:2023kyu}. This rather lower rate of the angular
momentum extraction from the BH can be attributed to the fact that in our case
the accretion flow stays in a mildly MAD state ($\phi_{\rm BH} \approx 20$).}

\subsection{Magnetic reversal} \label{sec:reversal}
\begin{figure}
\centering
\includegraphics[width=\linewidth]{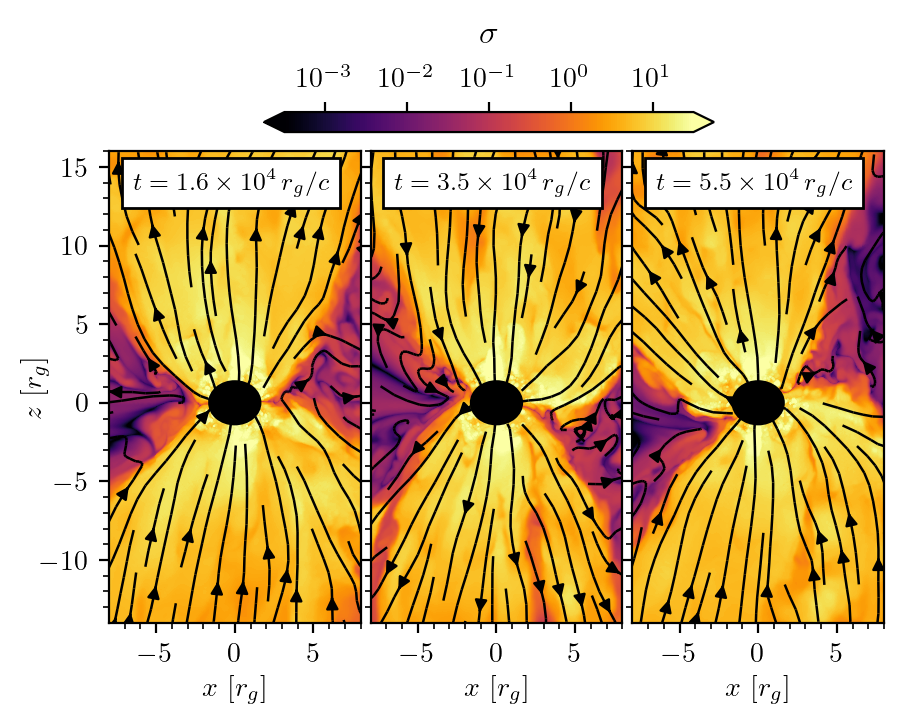}
\caption{Magnetic field reversal of the jets between three eruption epochs in
    the fiducial model {($\beta_{10}$-$\theta_{90}$-$R_{200}$)}. The
    relativistic magnetization $\sigma$ is shown in color and the in-plane
    components of the magnetic field are shown with black streamlines.}
\label{fig:reversal}
\end{figure}

Within the simulation time of the model ($t\leq 6\e{4}\,r_g/c$), the BH and the
accretion flow undergo two quiescent periods at $t=10\tau_a$ and $t = 24
\tau_a$. While all other quantities show a recurring pattern of rise and fall in
every eruption epoch, the vertical drag force $F^z$ (see the bottom line plot of
\reffig{fig:fiducial-timeseries}) shows a distinct behavior in that it flips its
sign during a quiescent period and maintains that opposite sign for the next
eruption epoch, before coming back to the original sign after another quiescent
period. 

We find that the polarity of the horizon magnetic fluxes and jets is reversed
during these quiescent periods, namely that the MAD state experiences a magnetic
reversal between eruption epochs; see \reffig{fig:reversal}. This reversal
behavior is observed from all models with a purely {horizontal} magnetic
field ($\theta_B = 90^\circ$) of the incoming fluid. Unfortunately, our
simulation is too short to conclusively decide whether the polarity selection
always alternates or is stochastic. We postpone a detailed discussion on the
origin of this phenomena to future studies.

\section{Dependence on accretion parameters}
\label{sec:results-parameter-studies}

In this section, we present a systematic survey of the accretion parameters. We
change the inclination angle $\theta_B$ between the incoming magnetic field and
the BH spin (\refsec{sec:results-inclination}), then explore a weakly magnetized
case with $\beta_\infty=100$ (\refsec{sec:results-high beta}), followed by
different incoming speeds of the fluid $v_\infty$ (\refsec{sec:results-wind
speed}). Rather than delving into the same level of detail as the previous
section, here we provide a broader overview on qualitative impacts of the
physical parameter chosen to be varied.

\subsection{Mixed magnetic fields}
\label{sec:results-inclination}

\begin{figure*}
\centering
\includegraphics[width=\linewidth]{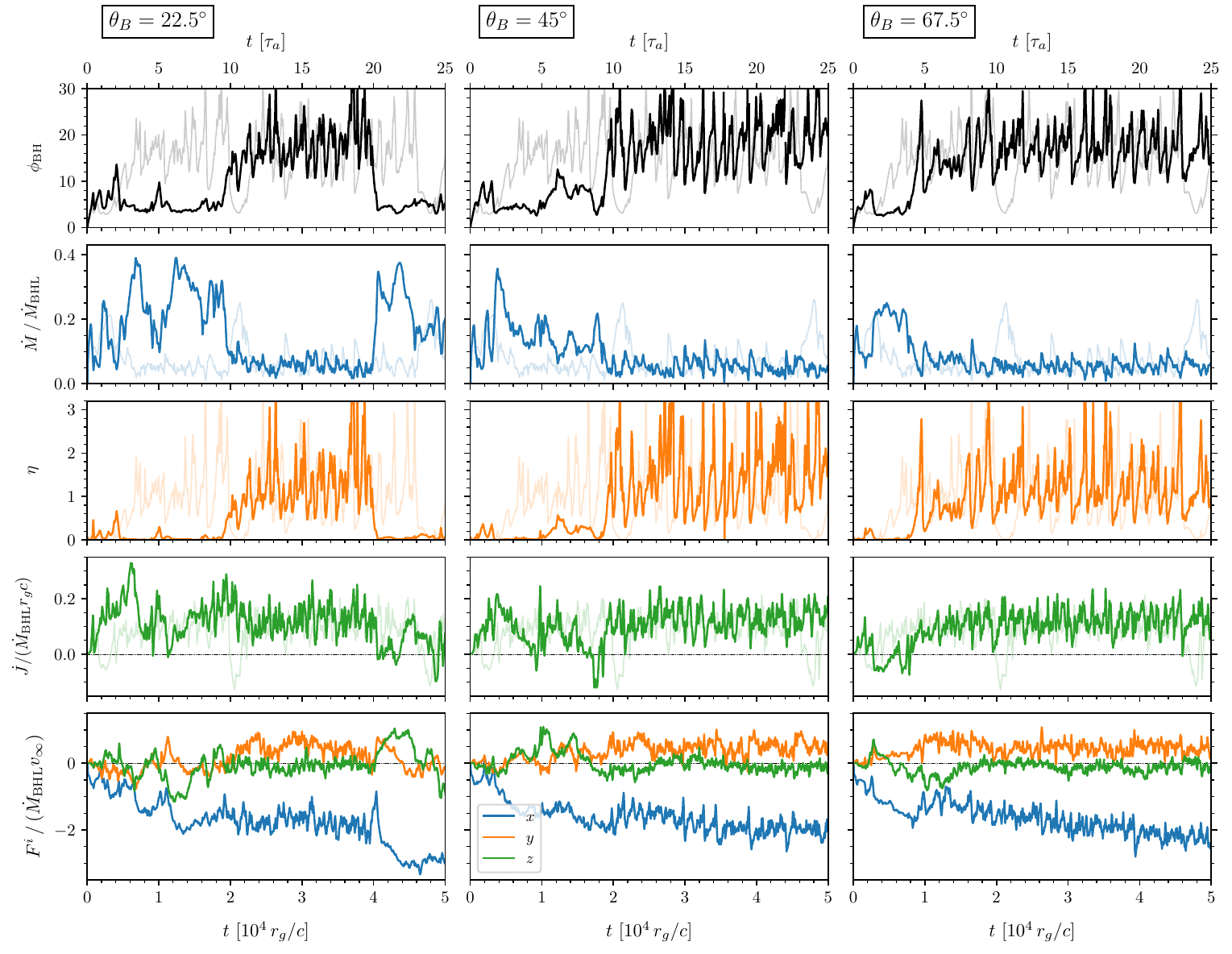}
\caption{Time evolution of physical quantities for three models with $\theta_B =
    23.5^\circ$, $45^\circ$ and $67.5^\circ$, all with $\beta=10$ and $R_a=200$.
    See Fig. \ref{fig:fiducial-timeseries} for a description of the quantities
    shown. {The result from the $\theta_B = 90^\circ$ (fiducial) model is
    overlayed with a transparent line.}}
\label{fig:inclination}
\end{figure*}

\begin{figure}
\centering
\includegraphics[width=0.9\linewidth]{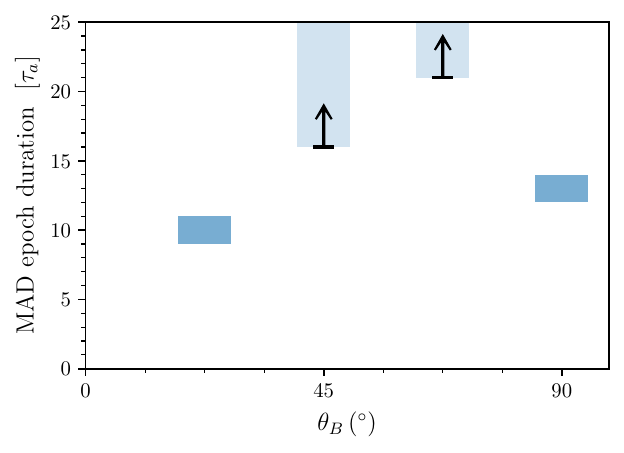}
\caption{Duration of a magnetically arrested accretion epoch in units of the
    accretion timescale $\tau_a = 2000 r_g/c$ for the four models with $R_a =
    200 r_g$ and $\beta_\infty = 10$.}
\label{fig:eruption-period}
\end{figure}

Keeping other parameters fixed as the fiducial setup, the inclination angle of
the initial magnetic field was varied to $\theta_B = 22.5^\circ, 45^\circ,
67.5^\circ$. The purpose of this experiment is to investigate how much the
{inclination} of the incoming magnetic field {relative to the BH spin}
affects jet launching and the time evolution of accretion flow.

In \reffig{fig:inclination}, we present the time series data from the three
simulations varying $\theta_B$. The result from the fiducial model
$\beta_{10}$-$\theta_{90}$-$R_{200}$ is overlayed with a transparent line in
each panels to highlight deviations. While the overall correlations between each
physical quantities is similar to that of the fiducial setup, we compile several
observations below.

\begin{enumerate}[label=(\roman*)]
\item The time it takes from the beginning of the simulation to the first
    successful jet launching and transition to the MAD state is longer for
    smaller inclination angle, namely when the incoming magnetic field has
    {more vertical} component. {Interestingly, the maximally misaligned}
    case ($\theta_{90}$) launches the jet the earliest.
\item Within the simulation time $t \leq 5\e{4} r_g/c$, both $\theta_B =
    45^\circ$ and $\theta_B = 67.5^\circ$ models did not show any quiescent
    period; the first eruption epoch continued throughout the final time. In
    contrast, the $\theta_B = 22.5^\circ$ model turned into a quiescent state
    after a single eruption epoch of a duration $10\tau_a$, and did not revive
    its jet activity until the end of the simulation, showing a quiescent period
    with the duration at least longer than $5 \tau_a$. The BH fails to establish
    a MAD state during this quiescent period, where its episodic launching of
    weak outflows ($\eta \lesssim 10\%$) into random directions appears to be
    very similar to what was observed from low angular momentum accretion flows
    in
    Refs.~\cite{2021MNRAS.504.6076R,Kwan:2022cnw,Lalakos:2023ean,Galishnikova:2024nsk}.
\item However, once the BH enters the eruption (MAD) state, values of
    $\phi_\text{BH}$, $\dot{M}$, $\eta$ and $\dot{J}$ are almost same as the
    $\theta_B=90^\circ$ case for all models. This implies that quasi-stationary
    properties of the MAD state in this setup do not depend on the large scale
    magnetic field geometry, which only determines the time period between
    eruption epochs.\footnote{{Ref.~\cite{Kaaz:2023b} adopts a similar
    parameter regime as in this work, apart from $\mach = 2.45$ and $\theta_B =
    0^\circ$. In an active (jet) state, the mass accretion rate
    $\dot{M}/\dot{M}_{\rm BHL}\approx 0.05$ and the energy outflow efficiency
    $1\lesssim \eta \lesssim 3$ observed in our simulation agree with those from
    \cite{Kaaz:2023b}, while the average value of $\phi_\text{BH}$ appears to be
    somewhat lower in our study.}}
    In \reffig{fig:eruption-period}, we plot the $\theta_B$ dependence of the
    duration of eruption epochs we could observe from our simulations, while we
    could only constrain lower bounds for $\theta_B = 45^\circ$ and $\theta_B =
    67.5^\circ$ cases. Our results indicate a non-monotonic behavior of the
    eruption epoch duration with the magnetic field orientation $\theta_B$, but
    an exact functional relationship between them is highly uncertain due to an
    insufficient number of data points and limited numerical resolution surveys.
\end{enumerate}

\begin{figure*}
\centering
\includegraphics[width=\linewidth]{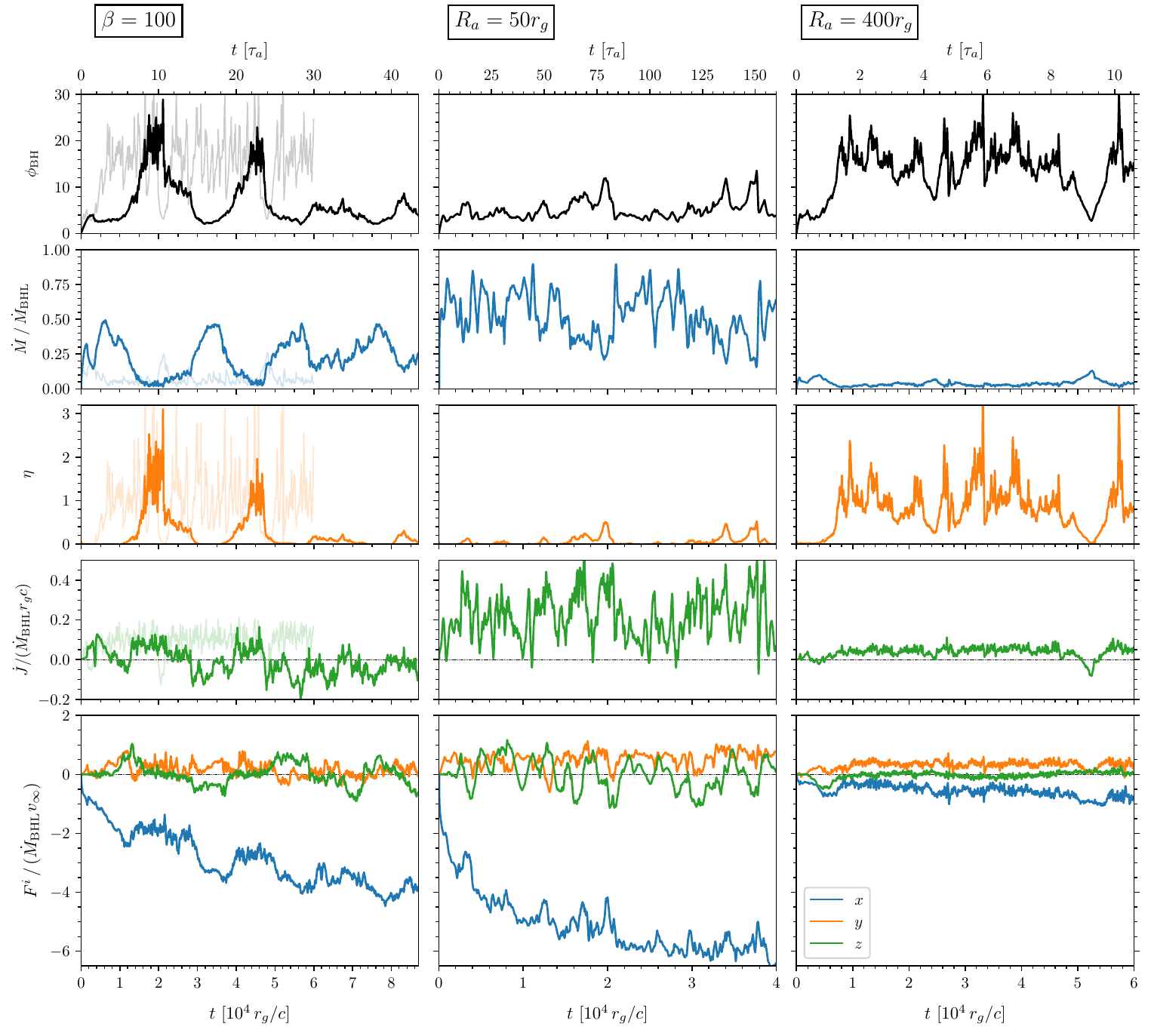}
\caption{Time series data for the $\beta_{100}$-$\theta_{90}$-$R_{200}$ (left),
    $\beta_{10}$-$\theta_{90}$-$R_{50}$ (center), and
    $\beta_{10}$-$\theta_{90}$-$R_{400}$ (right) models. See Fig.
    \ref{fig:fiducial-timeseries} for a description of the quantities shown.
    {We overlay the result from the fiducial
    ($\beta_{10}$-$\theta_{90}$-$R_{200}$) model with a transparent line only
    for the $\beta_{100}$-$\theta_{90}$-$R_{200}$ model (first column).}}
\label{fig:parameter-study-time-series}
\end{figure*}

\subsection{Lower magnetization}
\label{sec:results-high beta}

The first column of \reffig{fig:parameter-study-time-series} shows the time
series of accretion quantities for the model
$\beta_{100}$-$\theta_{90}$-$R_{200}$, which has ten times weaker magnetization
of the incoming fluid compared to the fiducial model. Several peculiar features
observed in this model compelled us to perform a longer time integration up to
$t = 8.7\e{4} r_g/c$. We outline those below.

The evolution of the horizon magnetic flux $\phi_\text{BH}$ roughly follows a
similar cycle. However, it undergoes more gradual increase and fall of
$\phi_\text{BH}$ between eruption and quiescent periods, exhibiting relatively
short duration of eruptions and a longer quiescent period. The BH cannot sustain
the MAD state long enough as the $\beta=10$ case and takes longer to revive jets
once it goes quiescent, which is consistent with the upstream wind providing a
smaller amount of magnetic flux per time. {These trends, along with an
increased $\dot{M}/\dot{M}_{\rm BHL}\approx 0.5$ in a quiescent period and
decreased average values of $\phi_{\rm BH}$, $\eta$ during an active period, is
similar to the results of Ref.~\cite{Kaaz:2023b} despite a different incoming
magnetic field geometry.}

{Another observation that can be made from the result, owing to a prolonged
duration of a quiescent period, is a reduced (or a reduced accumulation of)
dynamical drag $F^x$ over an eruption period. A strong, low-density bipolar jets
launched from the BH pushes the gas outward, dropping the average mass density
in the region trailing behind the BH \cite{Li:2019hfq}.}

The third eruption epoch ($28 \tau_a \lesssim t \lesssim 36 \tau_a$), though not
as conspicuous as the first two, ends up staying only in the mildly MAD regime
$\phi_\text{BH} \leq 10$. We observe the fourth round of rising magnetic fluxes
at $t=40 \tau_a$ during which horizon magnetic fluxes also stayed below
$\phi_\text{BH} = 10$. In brief, the system exhibits a periodic
eruption-quiescence cycle similar to what is observed in our fiducial model, but
with a decreasing level of activity over time. We caution that the decay we see
may be correlated with numerical resolution and scale separation between the BH
and the outer simulation domain boundary.

\subsection{Faster/slower incoming speed}
\label{sec:results-wind speed}

\begin{figure*}
\centering
\subfloat[$R_a = 50r_g$ (at $t=2\e{4}\,r_g/c$)]
{\includegraphics[width=0.48\linewidth]{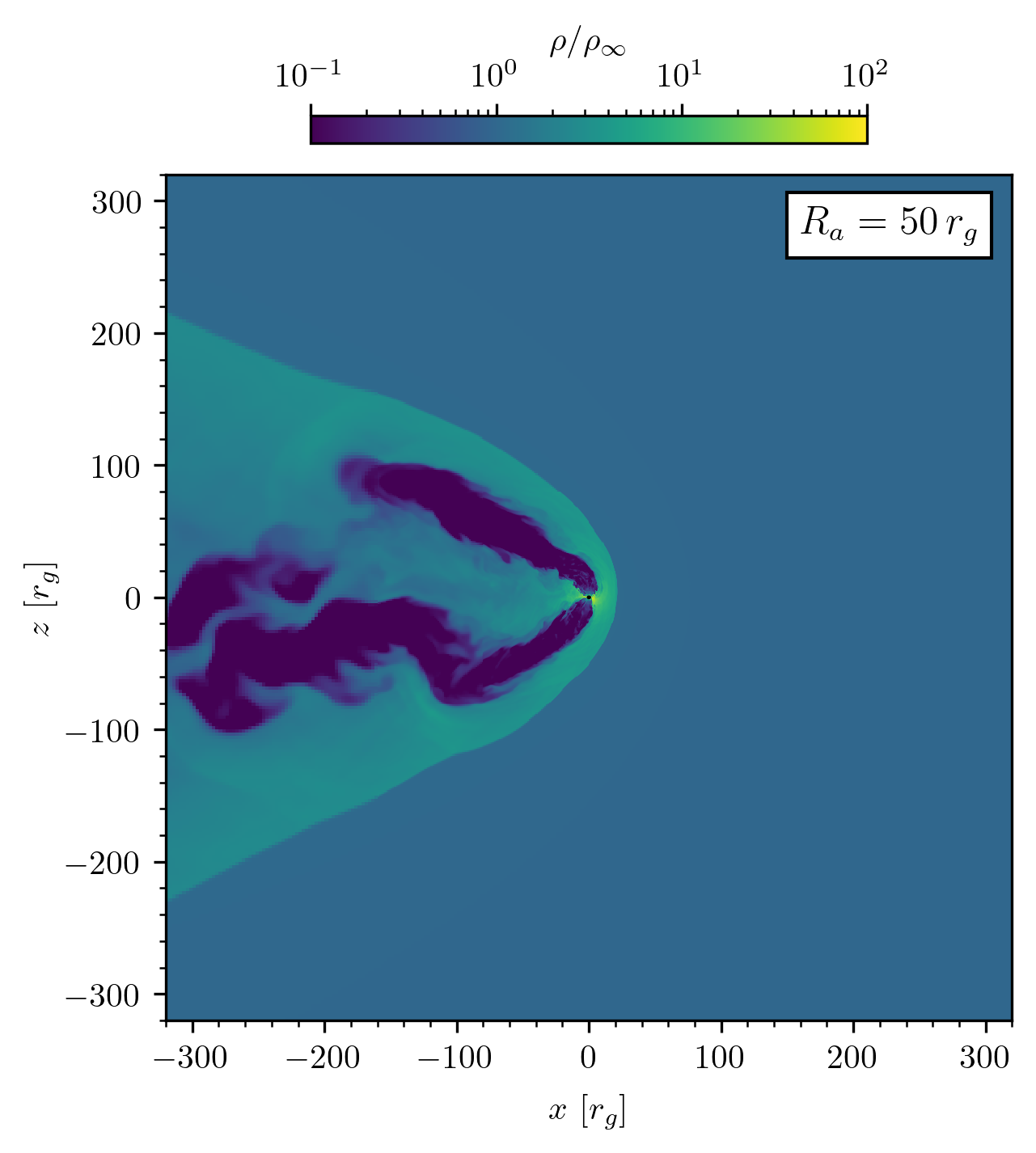}}
\hfill
\subfloat[$R_a = 400r_g$ (at $t=4\e{4}\,r_g/c$)]
{\includegraphics[width=0.48\linewidth]{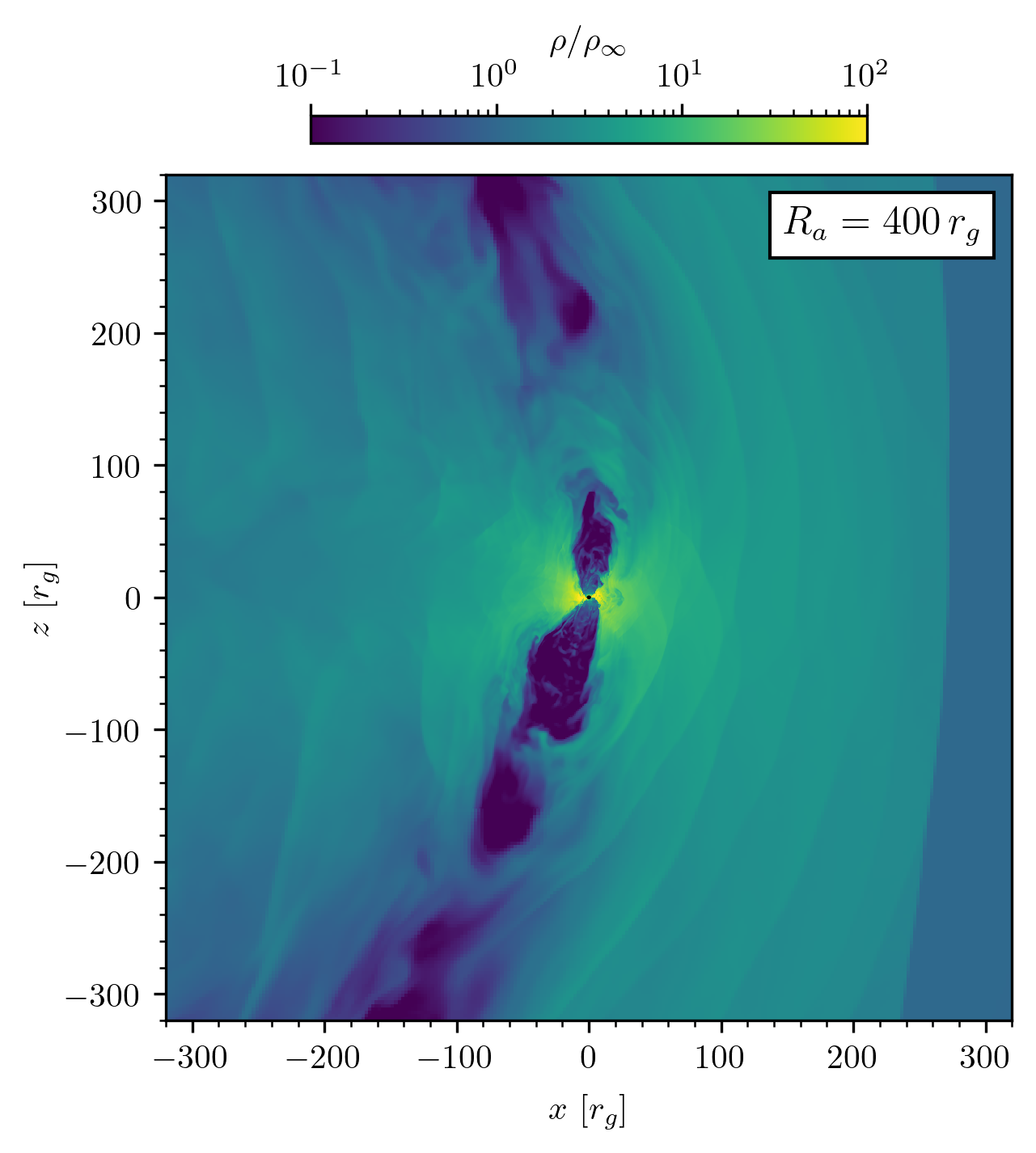}}
\caption{Distribution of the mass density $\rho$ in the meridional plane for the
    simulations $\beta_{10}$-$\theta_{90}$-$R_{50}$ (left) and
    $\beta_{10}$-$\theta_{90}$-$R_{400}$ (right). $R_a$ is the accretion radius,
    $r_g$ is the gravitational radius of the black hole, and $\rho_\infty$ is
    asymptotic mass density of the wind.}
\label{fig:speed}
\end{figure*}

Time evolution of the models with a faster ($R_a=50 r_g$, $v_\infty = 0.2c$) and
a slower ($R_a=400 r_g$, $v_\infty = 0.07c$) speed of the incoming fluid are
shown on the second and third columns in
\reffig{fig:parameter-study-time-series}. We show the mass density distribution
on the meridional plane for the two models in \reffig{fig:speed}, highlighting
differences in the shape and radius of the bow shock, as well as the bending
angle of outflows from the BH.

We first examine the case with a faster incoming speed ($R_a=50r_g$). The
outflow launched in the polar directions is choked by a strong ram pressure. The
BH exhibits sporadic flux eruptions and launches outflows intermittently, but
fails to maintain a continuous jet. While the horizon magnetic flux is kept
below $\phi_\text{BH} \leq 10$, the mass accretion rate shows a large
oscillation between 0.3\textendash0.9 $\dot{M}_\text{BHL}$, highly
anti-correlated with $\phi_\text{BH}$. While the dynamical time of the accretion
flow around the BH is the same independent of the wind speed, the replenishing
timescale of the flux changes with $R_a$. For faster wind speeds, the cross
section of the bow shock shrinks, with less mass being accreted on the BH,
though more relative to the changed reference rate, $\dot{M}_{\rm BHL}$. The BH
then preferentially accretes in a SANE regime, even if the magnetic properties
of the wind at large scales remain the same. This likely implies that questions
about flux accumulation on the black hole horizon in BHL accretion cannot be
separated from the effective speed of the black hole. We also observe that the
tangential drag reaches $f^x_\text{BHL}\approx 6$ by the end of the simulation;
recall that $f^x_\text{BHL}\approx 2.5$ in our fiducial model with $R_a = 200
r_g$. 

Next, we examine the model with a slower incoming speed of fluid ($R_a=400r_g$).
The accretion flow reaches the MAD state relatively early at $t \sim 1.5
\tau_a$. The shorter time for the flow to become magnetically arrested is
consistent with the scaling argument presented in \cite{Kaaz:2023b},
\begin{equation}
    \tau_\text{MAD} / \tau_a \propto R_a^{-3/4} ,
\end{equation}
while we note the exponent can be slightly different for {an inclined
magnetization of the inflow}. The accretion flow enters a quiescent period at $t
\sim 9 \tau_a$ and revives jets at $t\sim 10 \tau_a$, also showing a magnetic
reversal behavior. The overall time evolution of this model is qualitatively
very similar to the fiducial model discussed in \refsec{sec:results-fiducial}.
The dynamical friction measured by the end of the simulation was $f^x_\text{BHL}
\sim 1.0$. However, we expect a higher value in practice as the measured value
of the drag had not fully reached a steady state during our integration time. We
also observe that the magnitude of the vertical drag is reduced to
$|f^z_\text{BHL}| \sim 0.1$ and the correlation of its sign and the magnetic
polarity of the jet has become weaker compared to the fiducial model. The
vertical drag occasionally turns to a positive value during $2.5 \tau_a \leq t
\leq 4.5 \tau_a$ where the direction of magnetic field threading the BH and the
jet has been steadily kept to $+\hat{z}$ during the period. Considering that the
astrophysically realistic speed of the BH is still much beyond the lower limit
of $v_\infty$ explored in this study, it follows that the vertical drag $F^z$
may be effectively uncorrelated or only weakly correlated with the magnetic
polarity of jets in realistic situations.

\section{Discussion}
\label{sec:discussion}

\subsection{{Energy outflow}}

\begin{figure}
\centering
\includegraphics[width=\linewidth]{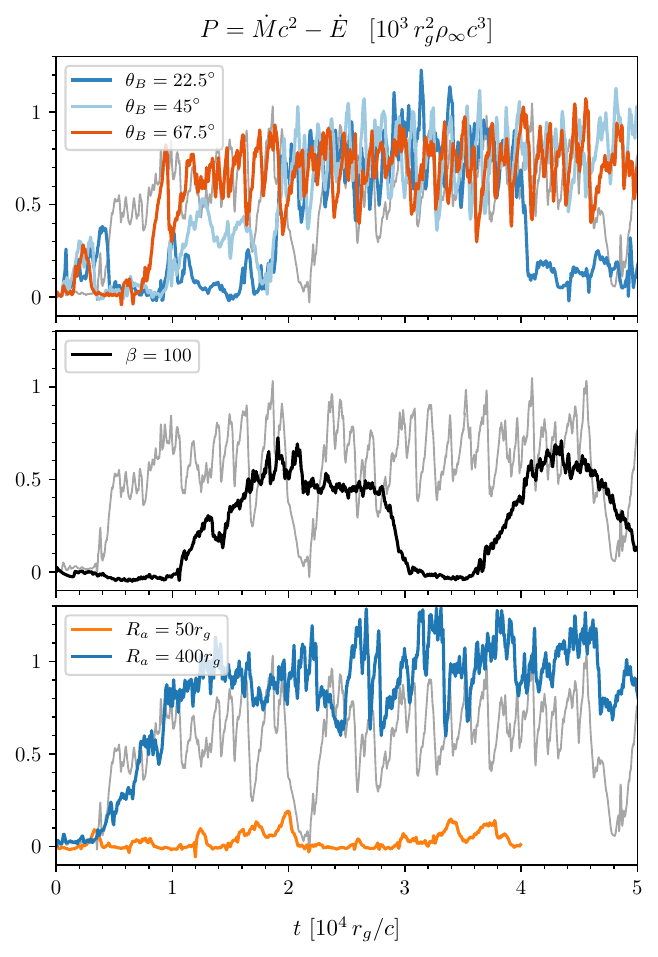}
\caption{{The energy outflow power $P = \dot{M}c^2 - \dot{E}$} from all
    simulations. The gray line shows the result from the fiducial model. For the
    scaled efficiency $\eta \dot{M} / \dot{M}_\text{BHL}$, see Appendix
    \ref{sec:conversion efficiency}}
\label{fig:jet-power}
\end{figure}

\reffig{fig:jet-power} compares the {net energy outflow} power ${P} =
\dot{M}c^2 - \dot{E}$ for all models. Once the accretion enters the MAD state,
the {energy outflow} power does not show much dependence on the inclination
angle $\theta_B$ of the incoming magnetic field. A slight decrease {in the
power} is observed when the magnetization of the incoming medium is lower, which
can be attributed to a reduced supply of magnetic energy to the BH per unit
time. Likewise, an increased {energy outflow} for a slower speed of incoming
fluid can be understood as that of a slowly moving BH with a large accretion
radius, leading to a higher mass accretion rate ($\dot{M}_\text{BHL} \propto
v_\infty^{-3}$), and consequently a higher rate of magnetic flux injection onto
BH, resulting in a more powerful {energy outflow} (and vice versa).

Our results indicate that among the accretion parameters varied in this study,
the {energy outflow} is most significantly influenced by the fluid speed
$v_\infty$ and to a lesser extent by the magnetization of the incoming matter
(note that the BH speed is $v_\infty$ varied by a factor of two at most, where
magnetization is set to be ten times weaker in the $\beta=100$ model). The
magnetic field inclination angle $\theta_B$ has a negligible impact on the
{outflow} power during eruption epochs. However, it largely effects the time
evolution and the intermittency of the jet activity, as previously discussed in
\refsec{sec:results-inclination}.

The finding from this comparative analysis, namely that the BH speed $v_\infty$
is a primary factor in determining the {outflow luminosity}, also aligns
with the fact that basic physical scales of the BHL accretion {e.g.,
Eq.~\eqref{eq:accretion radius}--\eqref{eq:bhl accretion rate}} possess a strong
dependence in $v_\infty$ with the highest power exponent.

A reference scale of the {energy outflow} power in physical units can be
written as
\begin{equation}
\begin{split}
    {P}
    & = \eta \left(\frac{\dot{M}}{\dot{M}_\text{BHL}}\right) \, \dot{M}_\text{BHL} c^2 \\    
    & = 1.0 \e{43} 
        \, \left(\frac{\eta \dot{M} / \dot{M}_\text{BHL}}{0.05}\right) 
        \left(\frac{M}{100 M_\odot}\right)^{2} \\
        & \hspace{3ex}
        \times
        \left(\frac{\rho_\infty}{10^{-10} \, \mathrm{g \, cm^{-3}}}\right)
        \left(\frac{v_\infty}{1000 \, \mathrm{km \, s^{-1}}}\right)^{-3} 
        \mathrm{erg\,s^{-1}} \, .
\end{split}
\end{equation}
The factor $\eta \dot{M} / \dot{M}_\text{BHL}$ corresponds to an effective
energy conversion efficiency with which the rest mass energy inflow
$\dot{M}_\text{BHL} c^2$ is converted into {a net energy outflow}. Our
baseline setup ($\beta_{10}$-$\theta_{90}$-$R_{200}$) shows the conversion
efficiency $\approx 0.05$ during an eruption epoch. See
\reffig{fig:conversion-efficiency} in Appendix \ref{sec:conversion efficiency}
for the values of $\eta \dot{M} / \dot{M}_\text{BHL}$ from all models.

\subsection{Magnus effect} \label{sec:magnus}

\begin{figure}
\centering
\includegraphics[width=\linewidth]{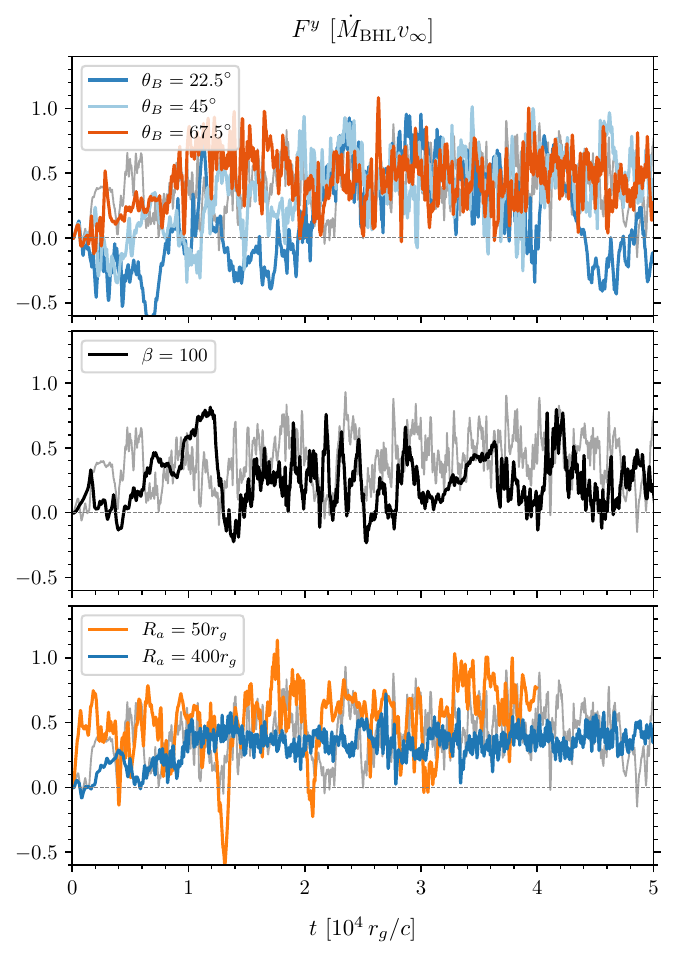}
\caption{The drag force to the transverse ($y$) direction (Magnus force) in all
    simulations. The gray line shows the result from the fiducial model. A
    negative value corresponds to the anti-Magnus effect.}
\label{fig:magnus}
\end{figure}

As can be seen from \reffig{fig:fiducial-slice-big}, the downstream flow
trailing the BH is deflected to the $-\hat{y}$ direction owing to the spin of
the BH and a resulting circulatory flow surrounding it. This can result in two
effects on the transverse drag force $F^y$: (1) the conservation of linear
momentum requires the BH to experience a reaction force to $+\hat{y}$ direction,
where (2) since a more amount of matter is deposited to $y<0$ region of the
downstream, a net gravitational pull from the flow is enhanced toward $-\hat{y}$
direction. These two effects are competing with each other, and the direction of
the net drag will be highly dependent on the nature of the accretion flow.

A nonzero transverse drag $F^y$ is the manifestation of the
(general-relativistic analog of the) Magnus effect, the phenomenon in the
classical fluid dynamics that a spinning body moving through a fluid
experiencing a drag force normal to both the direction of its motion and spin.
Although its physical origin is different, a similar effect is expected to be
present in general relativity when a spinning compact object is immersed in a
mass-energy current not aligned with its spin axis
\cite[{e.g.,}][]{Font:1999,Okawa:2014sxa,Costa_2018}.

However, the precise direction of this gravitational Magnus effect has been
under debate when considering non-hydrodynamical types of matter
\cite{Okawa:2014sxa,Cashen:2016neh,Costa_2018,Dyson:2024qrq,Wang:2024cej}. Ref.
\cite{Costa_2018} argues that the gravitational Magnus force consists of two
distinctive components (which they name as `Magnus' and `Weyl' respectively
therein) and especially the Weyl component can be highly dependent on the
specific scenario and boundary conditions of the physical system under
consideration. A recent fully relativistic analysis \cite{Dyson:2024qrq} has
shown that the drag is always anti-Magnus for a collisionless particle-like
matter field, where a wave-like scalar field shows a mixed behavior. A numerical
relativity simulation on the BHL accretion of scalar dark matter
\cite{Wang:2024cej} has reported an anti-Magnus effect.

Ref. \cite{Font:1999}, to the best of our knowledge, is the only numerical study
commenting on this phenomenon in the hydrodynamic regime, and suggested that the
enhanced pressure of the accretion flow on the counter rotating side of the Kerr
BH gives rise to the (pro-) Magnus effect, albeit without a quantitative
argument.

We report in this paper that our physical scenario---3D GRMHD BHL accretion onto
a spinning BH---yielded the positive sign of the Magnus force, analogous to the
one in classical fluid dynamics. \reffig{fig:magnus} collects and compares the
Magnus force $F^y$ from all models. For all the cases, the Magnus drag
maintained a positive value for the most of the simulation time. While the
gravitational drag force Eq.~\eqref{eq:gravitational drag} we compute is not a
fully general relativistic formula \cite[see
{e.g.,}][]{Costa_2018,Dyson:2024qrq,Wang:2024cej}, it is very unlikely to
change the direction of the Magnus effect we observe from simulations. We
mention that previous numerical studies on the gravitational Magnus effect in a
scalar field \cite{Okawa:2014sxa,Wang:2024cej} have been carried out in a
boosted metric, unlike our setup in which the black hole is fixed and the inflow
is imposed in terms of fluid velocity, potentially causing a quantitative
difference in the drag force.

It is also noteworthy that the magnitude of the Magnus force is not small, often
rising to a level comparable to the BHL drag force scale
$\dot{M}_\text{BHL}v_\infty$. Across all models, we quote a conservative overall
estimate that the Magnus force has been observed to be about 10\% of the
dynamical friction. In the circumstances that the initial linear momentum of the
BH has been substantially lost by the dynamical friction, its traveling
trajectory could have been largely deflected from its original direction of
motion.

Our results have implications for a number of astrophysical contexts.
In an extreme mass-ratio inspiral of binary black hole, if a secondary BH has
spin and is surrounded by gaseous medium, the gravitational Magnus force on the
secondary can alter its trajectory or excite an eccentricity to the orbit. The
resulting features in gravitational waves can be potentially detectable with
next generation gravitational wave detectors {e.g.,} LISA
\cite{LISAConsortiumWaveformWorkingGroup:2023arg}.
In the common envelope phase of a binary star, the drag force inside the gaseous
envelope is responsible for the orbital decay and expansion of the envelope
\cite{Livio1988,Taam:2000kj,Ivanova:2012vx}.
Our findings imply that the BH orbiting within the envelope could experience the
Magnus force when spiralling around the core of the companion star. Depending on
the relative orientation of the BH spin to its direction of motion, the Magnus
force can increase orbital eccentricity or induce precession of the orbital
plane. This may have a considerable impact on the evolution of the BH orbit over
long periods and the final configuration of the binary after the common envelope
phase. We note that in this context a wind profile varying in the transverse
direction needs to be taken into consideration as well, since the nonzero
gradient in mass density or velocity leads to a misaligned, rotated shock cone
geometry \cite{Cruz-Osorio:2016abh,Cruz-Osorio:2020dja,Lora-Clavijo:2015} which
in turn can significantly alter the direction of the total drag force.

\section{Conclusion}
\label{sec:conclusion}

We have conducted three-dimensional general-relativistic magnetohydrodynamic
simulations of Bondi-Hoyle-Lyttleton accretion onto a spinning black hole when
the magnetic field of the surrounding plasma is inclined with respect to the
spin of the BH. Our primary motivation was to investigate the dynamics of a BBH
merger remnant kicked into the disk of an active galactic nucleus, but owing to
the ubiquitous applicability of the Bondi-Hoyle-Lyttleton accretion problem, our
results are also applicable to broader astrophysical contexts. We summarize our
main findings below.

\begin{itemize}[leftmargin=3ex, labelsep=1ex]
\item The accumulation of magnetic flux onto the BH establishes a magnetically
    arrested state of the accretion flow and launches relativistic outflows to
    polar directions, which are bent toward the downstream at a larger radius.
    The accretion disk surrounding the BH extends up to a few tens of $r_g$ from
    the horizon, and is encompassed by a large-scale downstream flow in the
    shock cone.
\item Quasi-periodic magnetic flux eruptions from the BH launch pressure waves
    expanding the bow-shaped shock cone, and release strongly magnetized blobs
    (flux tubes)---which can potentially power flare-type electromagnetic
    transients (e.g., \cite{Hakobyan:2022alv,Zhdankin:2023wch}) ---along the
    equatorial plane into a narrow range of angles relative to the wind
    direction.
\item Anisotropic recoil from the magnetic flux eruptions drive a strong
    nutation on the accretion disk, often ripping its inner region off from the
    outer part. Nutation of the accretion disk is imprinted on the jet as its
    twisted morphology, {potentially} aiding the development of a kink
    instability \cite{Bromberg:2015wra}.
\item For a purely {horizontal} magnetized wind ($\theta_B = 90^\circ$), the
    system periodically undergoes a quiescent period with the duration $\gtrsim
    \tau_a$, during which jet is quenched and the accretion flow relaxes to the
    SANE state. Magnetic polarity inversion is observed during this period.
\item {With an increasing inclination of the magnetic field relative to the
    BH spin,} the jet is launched earlier. However, the {energy outflow}
    power and efficiency did not show significant differences once the system
    establishes a MAD state. The orientation of the incoming magnetic field
    appears to hardly affect steady-state properties of the jet, but determines
    the temporal behavior of its active and quiescent periods.
\item The model with a lower magnetization $\beta=100$ shows a more gradual
    evolution of $\phi_\text{BH}$ over active and quiescent epochs, as well as a
    slightly decreased {energy outflow}, which can be explained by a reduced
    supply of magnetic fluxes from the accretion flow.
\item When subjected to a faster wind speed, a decreased dynamical cross section
    and an increased ram pressure on the BH results in the suppression of jet
    launching. On the other hand, the model with a slower wind speed reached the
    MAD state earliest relative to the accretion timescale $\tau_a$, which is
    consistent with a qualitative argument made in \cite{Kaaz:2023b}.
    {Energy outflow} shows the strongest dependency in the wind speed among
    all parameters considered in this work. This strong dependence of the
    overall flow dynamics on the wind speed suggests that realistic values of
    $v_\infty$ will be a crucial element for improved models of GRMHD BHL
    accretion.
\item The gravitational Magnus effect is observed across all models, with the
    magnitude of a few tens of percents of the reference drag scale
    $\dot{M}_\text{BHL}v_\infty$. The direction of the Magnus force is the same
    as its classical aerodynamic counterpart.
\end{itemize}

Whereas the accretion radius $R_a$ adopted in this work has one of the largest
values in literature for the general relativistic BHL accretion, it is still
considerably far from a realistic condition. In the scenario of a kicked BBH
post-merger remnant, the recoil velocity is $\lesssim 200 \mathrm{\,km\,s^{-1}}$
for non-spinning binaries \cite{Gonzalez:2006md} where the spin effects can at
most enhance the recoil up to $\approx 4000 \mathrm{\,km\,s^{-1}}$ for superkick
configurations \cite{Campanelli:2007cga,Gonzalez:2007hi}. Newtonian studies
suggest that when a BBH is embedded in a gaseous environment, accretion makes
their orbital and spin axes aligned
\cite{Bogdanovic:2007hp,ColemanMiller:2013jrk}, suppressing the superkick
configuration. The maximum recoil for a spin-orbit aligned binary estimated from
numerical relativity simulations is $\approx 500 \mathrm{\,km\,s^{-1}}$
\cite{Herrmann:2007ac,Koppitz:2007ev,Healy:2014yta}. {Another physical system
that can harbor a fast wind accreting onto a BH is an X-ray binary with a
high-speed stellar wind; for example, Cygnus X-1 binary system features
$v_\infty \gtrsim 1000{\rm km/s}$ \cite{Davis1983,Gies:2008uh,Grinberg:2015iva}.
However, the orbital motion of the BH and a specific geometry of the wind
present in these systems might require a deviation from the conventional BHL
accretion setup.}

Our present study opens up a number of different avenues for future work, with
several questions still remaining to be answered. First, the {energy
outflow} power can be affected by parameters other than those we have considered
here: the magnitude of the BH spin, {the inclination of the magnetic field
relative to the wind velocity}, the spin-wind orientation
\cite{Gracia-Linares:2023yxw}, or hydrodynamic parameters such as the adiabatic
index and the Mach number which are known to strongly influence the stability of
the shock cone \cite{Foglizzo:2005in}. It would be also intriguing to
investigate the inflow with a nonzero net angular momentum
\cite{Cruz-Osorio:2016abh,Lora-Clavijo:2015}, which provides a more appropriate
scenario for the common envelope phase. The inclusion of radiative effects
\cite{Zanotti:2011mb} will also greatly change the dynamics for super-Eddington
accretion flows. Future investigations would help constructing a more detailed
physical picture of a black hole moving through a magnetized medium, with a
better bridging of the scale gaps between currently available numerical models
and realistic astrophysical scenarios.

Figures in this article were produced using Matplotlib \cite{matplotlib}, Numpy
\cite{numpy}, and Scipy \cite{scipy} packages.

\section*{Acknowledgments}

The authors are grateful to James Stone, Jacob Fields, and Hengrui Zhu for
technical support, and to Saavik Ford, James Fuller, Matthew Graham, {Yuri
Levin}, Barry McKernan, Nicholas Rui, and Alexander Tchekhovskoy for insightful
discussions.
{Y.K. is supported by the Sherman Fairchild Foundation and by NSF Grants No.
PHY-2309211, No. PHY-2309231, and No. OAC-2209656 at Caltech.}
E.R.M. gratefully acknowledges the hospitality of the Aspen Center for Physics,
which is supported by National Science Foundation Grant No. PHY-2210452. The
simulations were performed on DOE OLCF Summit under allocation AST198, and on
DOE NERSC Perlmutter under grant m4575. {This research used resources of the Oak
Ridge Leadership Computing Facility at the Oak Ridge National Laboratory, which
is supported by the Office of Science of the U.S. Department of Energy under
Contract No. DE-AC05-00OR22725, and resources of the National Energy Research
Scientific Computing Center, which is supported by the Office of Science of the
U.S. Department of Energy under Contract No. DE-AC02-05CH11231.}

\appendix
\newpage

\section{Extraction radius of density-related integral quantities}
\label{sec:r-dependence}

\begin{figure*}
\centering
\includegraphics[width=\linewidth]{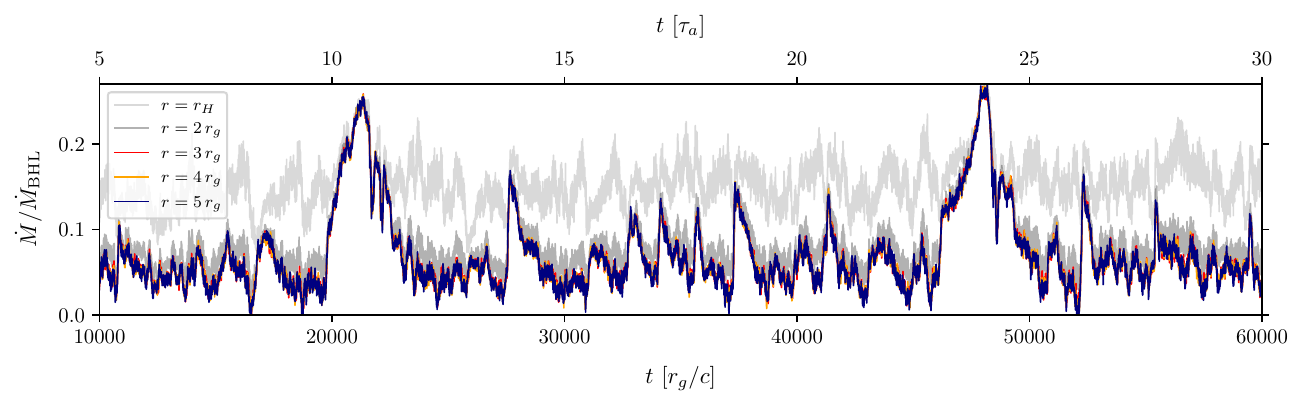}
\caption{Mass accretion rate extracted at different radii for the
    $\beta_{10}$-$\theta_{90}$-$R_{200}$ model. All data points are displayed
    without smoothing.}
\label{fig:mdot radial dependence}
\end{figure*}

\begin{figure*}
\centering
\includegraphics[width=\linewidth]{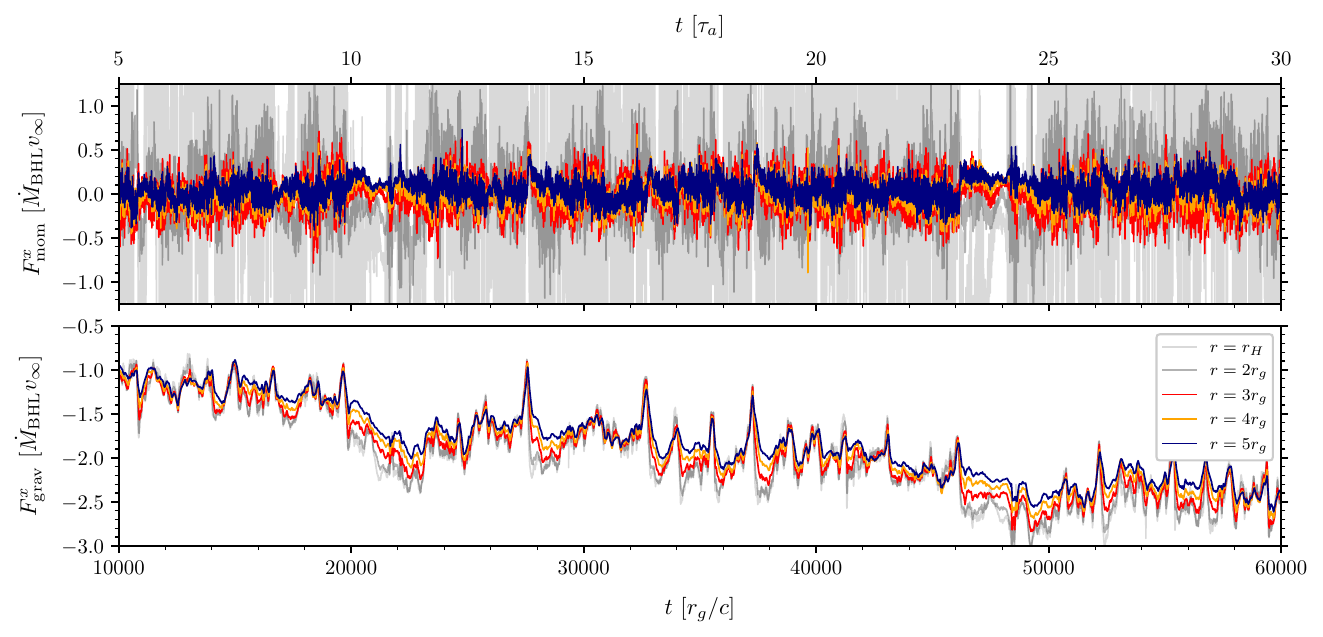}
\caption{The $\hat{x}$ component of the momentum drag (top panel) and
    gravitational drag (lower panel) computed with different radii from the
    $\beta_{10}$-$\theta_{90}$-$R_{200}$ model. All data points are displayed
    without smoothing.}
\label{fig:drag radial dependence}
\end{figure*}

As discussed in \refsec{sec:analysis}, the drift flooring algorithm artificially
injects the mass density floor to maintain the comoving magnetization $\sigma$
below an upper limit $\sigma_\text{max}$. This often induces spurious increases
in the fluid-related integral quantities when computed very close to the BH,
where the magnetization is very high.

In Figure~\ref{fig:mdot radial dependence}, we show the mass accretion rate
$\dot{M}$ integrated at different radii $r_i=\{r_H, 2r_g, 3r_g, 4r_g, 5r_g\}$
from our representative model ($\beta_{10}$-$\theta_{90}$-$R_{200}$) where $r_H
= r_g(1 + \sqrt{1-a^2/M^2})$ is the outer horizon radius. It can be clearly seen
that the artificial effects from numerical flooring becomes almost absent in
$r_i\geq 3r_g$.

\reffig{fig:drag radial dependence} shows, for the same set of radii, the
$\hat{x}$ component of the momentum drag $F^x_\text{mom}$ computed at the
spherical surface $r=r_i$ and the gravitational drag $F^x_\text{grav}$
integrated over the whole computational domain except the spherical volume $r <
r_i$. The momentum drag settles down to near zero for $r_i\geq 3r_g$. The
magnitude of the gravitational drag monotonically decreases for a larger $r_i$
since the region $r< r_i$ is simply excluded from the volume integral. The
differences between them are not significant, indicating that the gravitational
drag is mostly contributed from $r \geq 5r_g$.

\section{Unit conversion}\label{app:units}

From simulation results, physical values can be recovered as
\begin{align}
    x^i & = r_g \, \hat{x}^i \\[1ex]
    t & = (r_g/c) \, \hat{t} \\[1ex]
    \rho & = \rho_\infty \, \hat{\rho} \\[1ex]
    \dot{M} & = (r_g^2 \rho_\infty c) \hat{\dot{M}} \\[1ex] 
    \dot{E} & = (r_g^2 \rho_\infty c^3) \hat{\dot{E}} \\[1ex]
    \dot{J} & = (r_g^3 \rho_\infty c^2) \hat{\dot{J}} \\[1ex]
    {B^i} & = {(\rho_\infty^{1/2} c) \hat{B}^i} \\[1ex]
    \Phi_\text{BH} & = (r_g^2 \rho_\infty^{1/2} c) \hat{\Phi}_\text{BH} \\[1ex]
    F^i & = (r_g^2 \rho_\infty c^2) \hat{F^i}
\end{align}
where hat variables are the results in the (scale-free) code unit.

\section{Energy conversion efficiency} \label{sec:conversion efficiency}

\begin{figure}
\centering
\includegraphics[width=\linewidth]{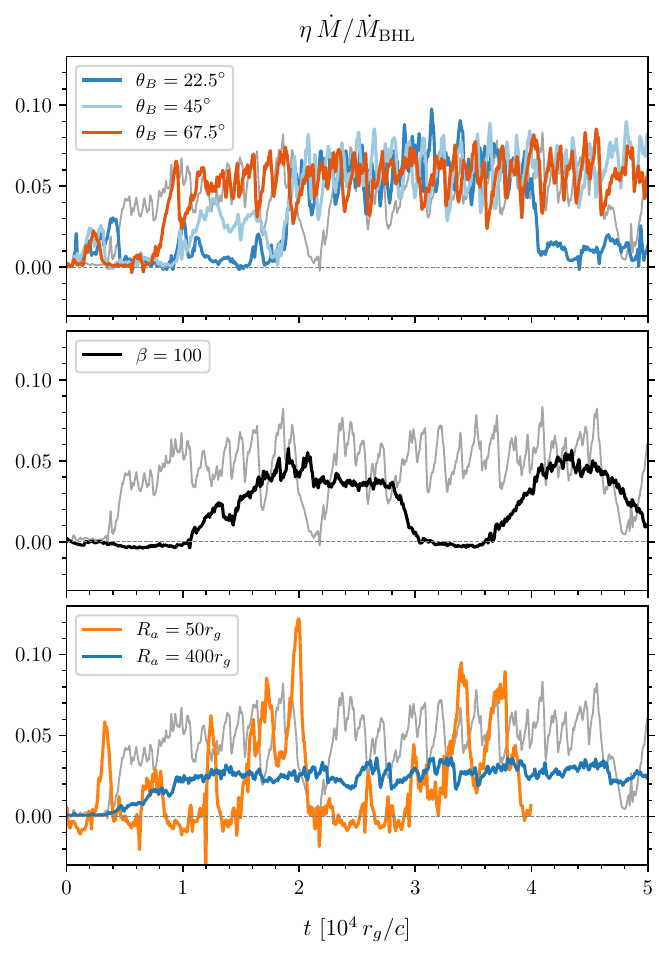}
\caption{{The energy conversion efficiency $\eta \dot{M}/\dot{M}_{\rm BHL}$ from
    all simulations. The gray line shows the result from the fiducial model
    $\beta_{10}$-$\theta_{90}$-$R_{200}$.}}
\label{fig:conversion-efficiency}
\end{figure}

{In Fig. \ref{fig:conversion-efficiency}, we show the energy conversion
efficiency $\eta \dot{M}/\dot{M}_{\rm BHL}$ from all simulations.}

\bibliography{references}
\end{document}